\documentclass[aps, prb, a4paper, reprint, 10pt, showpacs, twocolumn, groupedaddress, longbibliography]{revtex4-2}

\usepackage{graphicx}
\usepackage{amsmath}
\usepackage{amssymb}
\usepackage{bm}
\usepackage{natbib}
\usepackage{subcaption}
\usepackage{inputenc}

\begin{document}
  
\title{Supercurrent in the presence of direct transmission and a resonant localized state}
\author{Hristo Barakov} \author{Yuli Nazarov}
\affiliation{Kavli Institute of Nanoscience, Delft University of Technology, 2628 CJ Delft, The Netherlands} 

\begin{abstract}
Inspired by recent experimental findings that will be presented elsewhere, we formulate and investigate a model of a superconducting junction that combines the electron propagation in a quantum channel with an arbitrary transmission, and that through a localized state. Interesting situation occurs if the energy of the localized state is close to Fermi level, that is, the state is in resonant tunnelling regime. Since this energy is affected by the gate voltage, we expect a drastic modification of transport properties of the junction in a narrow interval of the gate voltages where the energy distance to Fermi level is of the order of $\Gamma, \Delta$, $\Gamma$ being the energy broadening of the localized state, $\Delta$ being the superconducting energy gap.

We consider the model neglecting the interaction in the localized state, as well as accounting for the interaction in a simplistic mean-field approach where it manifests itself as a spin-splitting. This spin splitting is also contributed by external magnetic field. We also take into account the spin-orbit interaction that can be significant in realistic experimental circumstances.

In normal state, we find that the model may describe both peak and dip in the transmission at resonant gate voltage. Spin splitting splits the positions of these peculiarities. Fano interference of the transmission amplitudes results in an asymmetric shape of the peaks/dips. In superconducting state, the spin splitting results in a complex dependence of the superconducting current on the superconducting phase. In several cases, this is manifested as a pair of $0 - \pi$ transitions in the narrow interval of gate voltages.

The article in the present form is not intended for a journal submission.

\end{abstract}

\maketitle
\thispagestyle{empty}

\section{Scope, style and structure of the article}
In its present form, this article is not intended for a submission to a journal. We believe that the theory developed here is worth a journal publication only together with the account of experimental activities, and full comparison of experimental and theoretical findings. This publication is in preparation.

We also find the model to be of significant general interest for current research in superconducting nanostructures. Despite the basic simplicity, the derivation of the model and elaboration on concrete results invokes a big number of technical details which are not normally given in a journal publication. So we chose to share our results in the present form that gives a full account of these technical details.

As a matter of style, this article does not include a usual introductory part and is not accompanied by proper references.  

The structure of the article is as follows. In Section \ref{sec:experiment}, we give a short summary of our impression of the experimental results. We explain motivation of the model and list its key ingredients in Section \ref{sec:motivation}. The Hamiltonian formulation is given in Section \ref{sec:ham}.
In Section \ref{sec:many} we derive the Landauer description of normal electron transport for an arbitrary number of dots and leads. We specify to two-dot, two-lead model in Section \ref{sec:twodots} where we perform the necessary derivations to adjust the model to the situation at hand for the case of normal transport.  The illustrative normal transport examples are given in Section \ref{sec:normaltransportexamples}. We turn to theoretical description of superconducting transport in Section \ref{sec:suptrans} and describe our numerical methods in Section 
\ref{sec:numerics}. The most important Section \ref{sec:supexamples} prodives several examples of superconducting transport.
We conclude in Section \ref{sec:conclusions}.

\section{Short summary of experimental observations}
\label{sec:experiment}
Let us shortly present the essence of experimental findings that inspired us to elaborate on the model. These experiments have been performed by V. Levajac, J. Y. Wang, L.P. Kouwehnoven, and other members of their team at QuTech, Delft University of Technology. The proper account of the experiments will be published elsewhere. Here we present our personal (theoretical) impression of the results.

\begin{figure}[tb]
\begin{center}
\includegraphics[width=0.92\columnwidth]{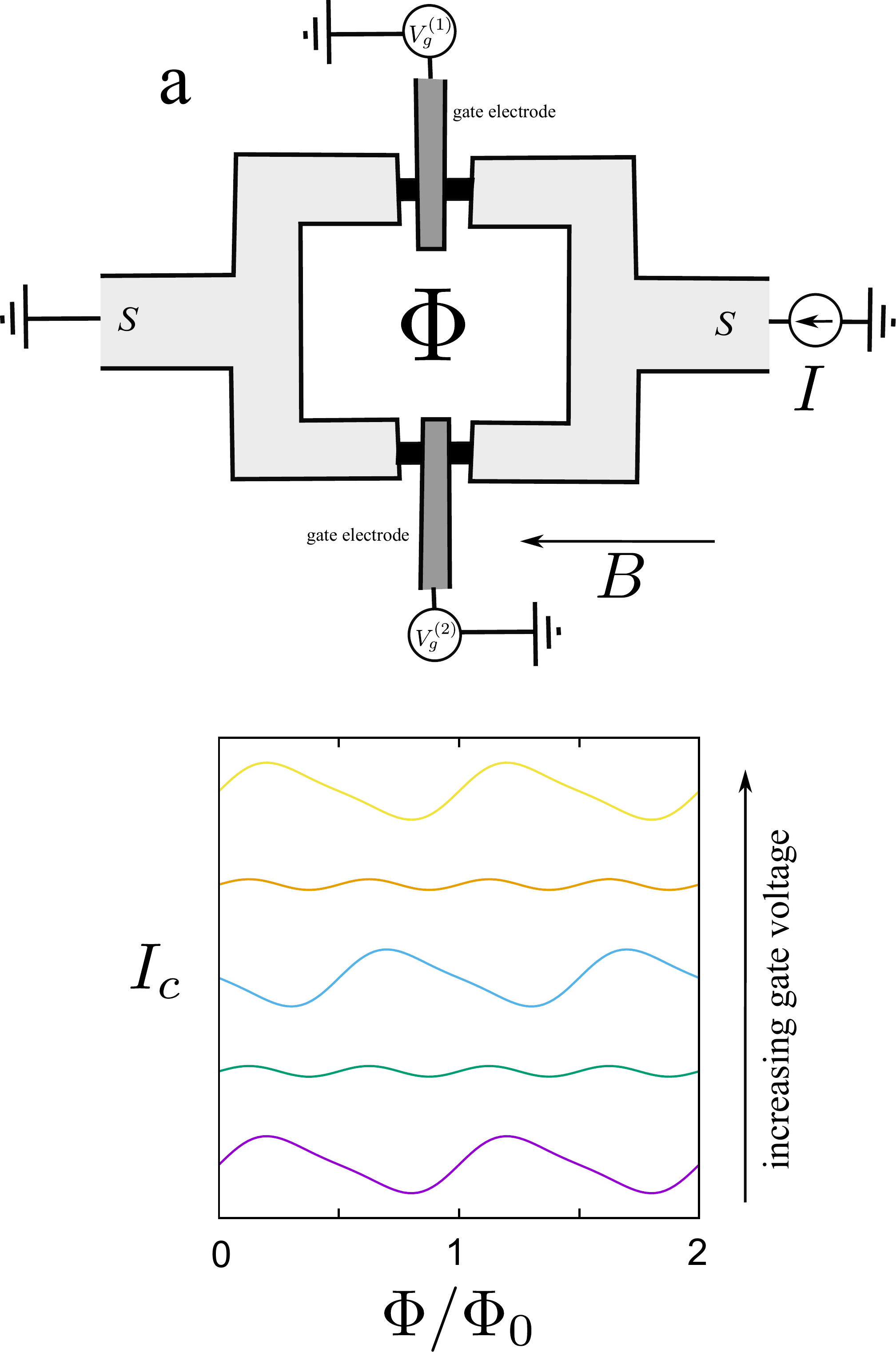}
\end{center}
\caption{a. Scheme of the setup (not in scale). Two semiconducting wires (black) are covered with a superconducting film (light grey) forming two Josephson junctions in a SQUID loop. The wires are affected by the voltages applied to the gate electrodes (dark grey).The loop is penetrated by magnetic flux $\Phi$. The parallel magnetic field $B$ may be applied. b. An intriguing observation: a pair of $0-\pi$ transitions in a narrow interval of a gate voltage. The curves give the dependence of critical current on the flux in the loop for a set of increasing gate voltages and are offset for clarity.}
\label{fig:experiment}
\end{figure}

The setup involves two superconducting junctions made by covering a semiconducting nanowire with superconducting electrodes. The junctions are enclosed in a SQUID loop that enables to characterize the dependence of the currents in the junctions on the superconducting phases changing the magnetic flux in the loop. A substantial magnetic field can be also applied in the plane of the substrate. There are gate electrodes affecting the junctions separately. The measurement is a simple voltage measurement at a given current bias. (Fig. \ref{fig:experiment} a). From this, one can inherit the critical current of two junctions in parallel. Another parameter that can be varied in this experiment is the magnetic field in the plane of substrate, parallel field. 

Naturally, the supercurrents vary smoothly upon changing the gate voltages at various magnetic fields. This is explained by depletion/addition of electron density to the junction that closes/opens the transport channels and modulates their transparency. The conductances of the junctions are several $G_Q\equiv e^2/\pi \hbar $ suggesting 1-2 open transport channels. An unusual observation the experimentalists share with us is a sharp dependence of the supercurrent upon changing one of the gate voltages in a narrow interval. In this interval, the change of the electron energies induced by the gate voltage is of the order of $1 {\rm meV}$, that is comparable with the value of the superconducting gap and Zeeman energy coming from the parallel field.

Some data can be interpreted as two close $0-\pi$ transitions in this narrow interval of gate voltages. In an idealized case (which is not necessary an experimental one) where the supercurrents through the junctions differ much in the magnitude, the $I-\phi$ dependence of the Josephson current in the junction with smaller current can be directly seen in the dependence of the critical current on the flux $\Phi$ in the SQID loop. If one changes the gate voltage controlling the smallest junction, the observation could be then summarized as follows
(Fig. \ref{fig:experiment} b): i. the positions of supercurrent minima are close to $\Phi_0/4+ n \Phi_0$ indicating the minimum of Josephson energy at $\phi=0$ ii. $\pi$-shifted dependence in the middle of the interval indicating the minimum of Josephson energy at $\phi=\pi$ iii. Double periodicity of the current at the borders of the interval.

Such pairs of close $0-\pi$ transitions occur may occur several times at different gate voltage settings. The widths of the interval increases upon increasing the parallel magnetic field. Sometimes the transitions merge and disappear at small magnetic field. Sometimes the effect persists even at zero field. 

\section{The motivation and essence of the model}
\label{sec:motivation}
The sharp dependence on the gate voltage in a narrow interval suggest that a localized state is involved. The gate voltage shifts its energy level with respect to Fermi energy. Beyond the interval, the state is either empty or occupied  and hardly participates in transport, either normal or superconducting. In the interval, resonant transport occurs via the state. The width of the interval is set by either $\Gamma$, the width of the level due to escape to the leads, or $\Delta$, the superconducting energy gap.

There are known mechanisms of $0-\pi$ transitions involving a localized state. First one is due to spin splitting of Andreev states in magnetic field. If the splitting is of the order of $\Delta$, the curvature of Andreev levels at zero phase may be inverted, and the Josephson energy achieves minimum at $\phi=\pi$ rather than zero. If interaction in the localized state is essential, the state is single-occupied in an interval of the gate voltage, and the minimum of Josephson energy may be at $\phi=\pi$ in this interval (Contrary to a popular belief, this is not always true for a single-occupied state). These mechanisms are not mutually exclusive but rather related: in a mean-field approximation, the interaction may be described as a spin splitting, and the field-induced splitting leads to single occupation if the chemical potential is between the split levels. This provides extra motivation to explain the experimental observation with a localized state.

However, the situation is obviously more complex than just the transport through a localized state. At least a single transport channel is open when the localized state becomes resonant. A very simplistic model would be independent parallel transport in the localized state and in the channel. This model, however, is not flexible enough to fit the experimental data. We need to take into account interference of transmissions through the channel and the resonant states. 

\begin{figure}[tb]
\begin{center}
\includegraphics[width=0.92\columnwidth]{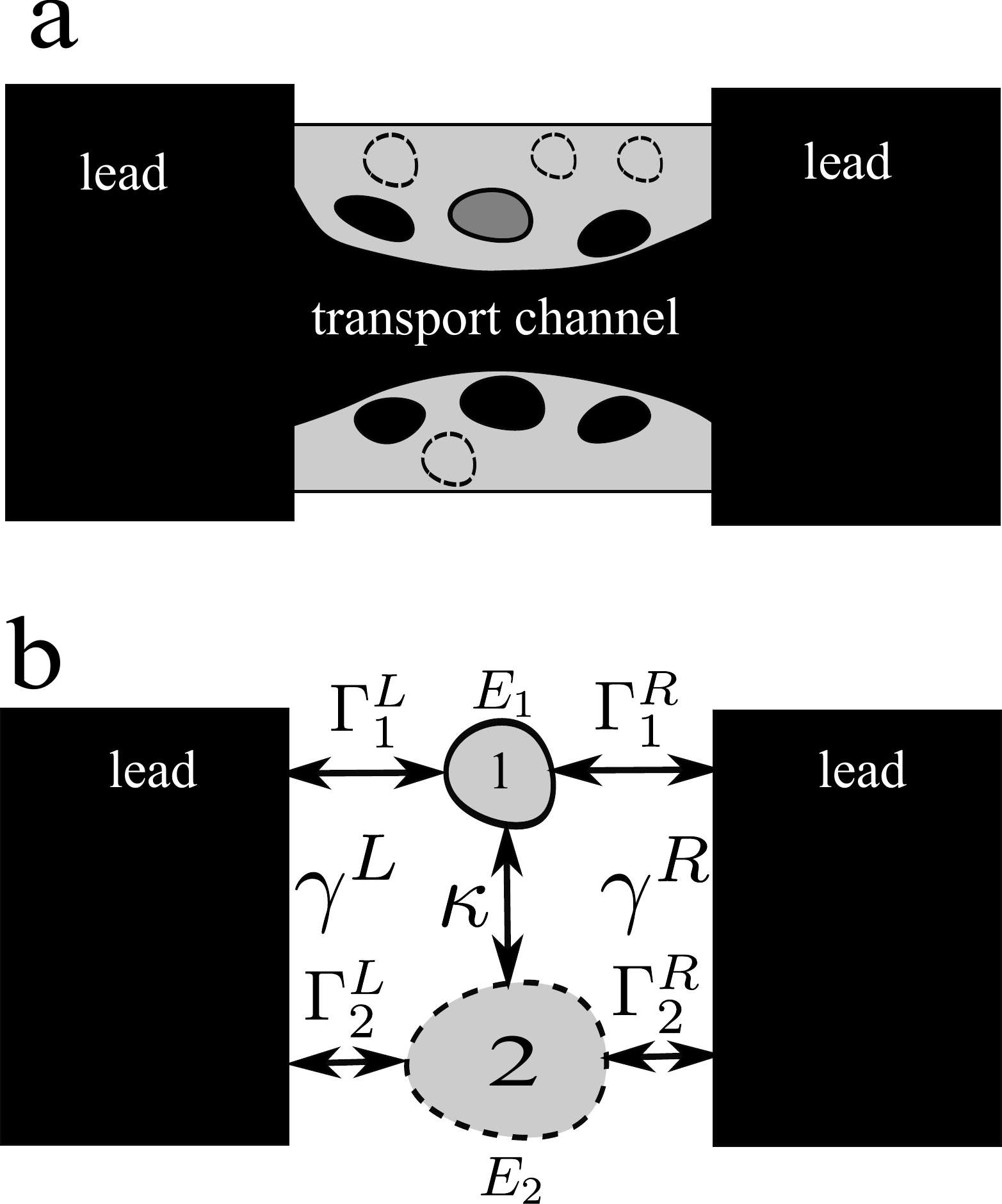}
\end{center}
\caption{a. Cross-section of the wire. Geometry of electron distribution in the wire and the leads. The filled states are given in black. The electron density is depleted near the wire surface. Random potential minima are either filled (small black regions) or empty (regions with dashed boundaries). The resonant state is given in dark grey.  b. Essential parameters of the two-dot model in use. The second dot is only use to simulate a transport channel with a transmission not depending on energy, so that 
the tunnel rates $\Gamma_{1,2}^{R,L}$ and the positions of the dot energy levels with respect to Fermi level, $E_{1,2}$, satisfy $\Gamma_2^{R,L}, E_2 \gg \Delta \simeq \Gamma_1^{R,L}, E_1$. There is also tunneling $\kappa$ between the dots, and the tunneling rates $\gamma_{R,L}$ that cannot be ascribed to a certain dot.}
\label{fig:model}
\end{figure}

A motivation for this also comes from the presumed geometry of electron distribution in the nanowire: the localized states are most likely appear in random potential minima of a nanowire part where the density is depleted. (Fig. \ref{fig:model} a)
Electron tunneling from these minima may proceed to the leads as well as to the transport channel. There may be many such minima that are subsequently filled upon changing the gate voltage giving rise to many localized states. We note that only the states with the escape rate $\Gamma \simeq \Delta$ may be responsible for the observed peculiarity. Those with $\Gamma \gg \Delta$  modify the transport smootly at the energy scale $\Gamma$, so at the energy scale $\Delta$ that is relevant for superconducting transport would only cause the renormalization of the transmission coefficients of the transport channels. Those with $\Gamma \ll \Delta$ modify the transmission only in a narrow energy interval: this would not give rise to Andreev states that require significant transmission at two opposite energies (for electrons and holes). 

To formulate a practical model encompassing the channel and the localized state, we note that the transport channel can be conveniently modelled with a localized state as well, provided the escape rate of this state $\Gamma_2$ by far exceeds $\Delta$. So we elaborate the model that encompasses two localized states, or dots, that are connected to two leads by tunnelling. Before writing any Hamiltonians, let us list most important parameters of the model (see Fig. \ref{fig:model} b). Two dots are at energy levels $E_{1,2}$ and are connected to left and right lead by tunneling with the rates $\Gamma_{1,2}^{L,R}$. There is a direct tunnel coupling $\kappa$ between the dots. Important non-trivial element are tunneling rates $\gamma_{L,R}$ that can not be ascribed to a certain dot but are requied to describe tunneling of a superposition state of to dots. Since the second dot is here only to model a channel, the model only makes sense under assumption $\Gamma^{L,R}_2 \gg \Gamma^{L,R}_1, E_1$. Owing to this, we can neglect the influence  of the gate voltage and magnetic field on $E_2$. The parameters $\kappa,\gamma$ are at intermediate scale, $\kappa, \gamma \simeq \sqrt{\Gamma_1 \Gamma_2}$. 

The most important interaction in this model is the on-site interaction in the localized state. It would be tempting to neglect this interaction, since we cannot treat it exactly. Besides, the localized state is near the transport channel so the interaction should be strongly suppressed by screening. However, the $0-\pi$ transition pairs are sometimes observed at zero magnetic field, this suggest that interaction should play a role. We compromise by treating the on-site interaction in a simple mean-field approach. 

The bandstructure of the semiconductor material of the wire provides strong spin-orbit interaction that we also include to the model. The coefficients $\kappa, \gamma_{L,R}$ therefore posses the corresponding spin structure.

\section{Hamiltonians}
\label{sec:ham}
In this Section, we give the Hamiltonians of the constituents of our model.
\subsection{The single dot}
We start with a dot Hamiltonian. It involves on-site annihilation operators $\hat{d}_\alpha$,
$\alpha$ being the spin index, and reads
\begin{equation}
\hat{H}_D = \hat{d}^\dagger_\alpha H_{\alpha \beta}\hat{d}_\beta + U \hat{n}_\uparrow \hat{n}_\downarrow
\end{equation}
$\hat{n}_\alpha = \hat{d}^\dagger_\alpha \hat{d}_\alpha$. The single-particle Hamiltonian reads
$$
\check{H} = E + {\bf B}\cdot \boldsymbol{\sigma}
$$
${\bf B}$ being the magnetic field, $\boldsymbol{\sigma}$ being the vector of Pauli matrices.

Importantly, we treat the interaction in the mean-field approximation. If there is a natural quantization axis (that can be absent in the presence of SO interaction in the coupling to the leads), the mean field gives the following  additions to the single-particle Hamiltonian, 
\begin{equation}
H_{\uparrow\uparrow} = U \langle \hat{n}_\downarrow\rangle;  H_{\downarrow\downarrow} = U \langle \hat{n}_\uparrow\rangle.
\end{equation}
In general situation, 
\begin{equation}
H_{\alpha\beta}= U\left(\delta_{\alpha\beta}\langle \hat{N}\rangle - \langle \hat{d}^\dagger_\alpha \hat{d}_\beta\rangle \right)
\end{equation}
The advantage of this mean-field scheme is that it delivers exact results in the absence of tunnel coupling. In particular, at zero magnetic field the
ground state corresponds to single occupation of the dot in the interval
$U>E-\mu>0$. At the ends of the interval, sharp transitions bring the dot to the states of zero and double occupation. The scheme is approximate in the presence of tunnel coupling, yet we use it for the lack of better general approach to interaction. 

\subsection{The leads}

We introduce annihilation operators in the leads 
$\hat{c}_{k,\alpha}$ where $k$ labels the states of quasi-continuous spectrum in the leads. The states $k$ are distributed over the leads, those are labelled with $a$. We assume the states $k$ are invariant with respect to time inversion.

The leads are described by the usual BSC Hamiltonian
\begin{equation}
\hat{H}_{\rm leads} = \sum_k \xi_k \hat{c}^\dagger_{k,\alpha}\hat{c}_{k,\alpha} + \sum_a \sum_{k \in a} \left(\Delta^*_a \hat{c}_{k,\uparrow} \hat{c}_{k,\downarrow} + {\rm h.c} \right)
\end{equation}
$\xi_k$ are the energies of the corresponding states. The superconducting order parameter $\Delta_a$ is different in different leads. To describe normal leads, we just put $\Delta_a=0$.

\subsection{Tunnel coupling}
The tunnel coupling to the states is described by the following Hamiltonian
\begin{equation}
\hat{H}_T = \sum_k \hat{c}^\dagger_{k,\alpha} t^{k}_{\alpha\beta} \hat{d}_\beta + {\rm h.c}
\end{equation}
For time-reversible case, the tunnel amplitudes are given by
\begin{equation}
\check{t} = t_k + i {\bf t}_k \cdot \boldsymbol{\sigma}
\end{equation}
with real $t_k,{\bf t}_k$. Of course, the multitude of  tunneling amplitudes comes to the answers only in a handful of parameters. One of such parameters is the decay rate from the dot to the continuous spectrum of the lead $a$,
\begin{equation}
\Gamma_a(\epsilon) = 2\pi \sum_{k \in a} \left(|t_k|^2 + |{\bf t}_k|^2\right) \delta(\xi_k - \epsilon)
\end{equation}
One can disregard the dependence of the rates on the energy $\epsilon$.

 
\section{Normal transport for many dots}
\label{sec:many}
In this Section, we will derive the currents in the nanostructure assuming the leads are normal and are kept at different filling facts. We do this derivation for an arbitrary number of the leads and dots, and later specify this for two dots and two terminals. Let us consider the following Hamiltonian where we do not specify spin or dot structure
\begin{equation}
\hat{H} = \sum_k \xi_k \hat{c}_k^\dagger c_{k} + \hat{d}^\dagger_\alpha H_{\alpha \beta}\hat{d}_\beta + \sum_k (\hat{c}^\dagger_{k} t^{k \beta } \hat{d}_\beta + h.c )
\end{equation}
The Heisenberg equations read
\begin{eqnarray}
i \dot{\hat{c}}_k &=& \xi_k \hat{c}_k + t_{k\alpha} \hat{d}_\alpha \\
i \dot{\hat{d}}_\alpha &=& H_{\alpha\beta} \hat{d}_\beta +t^*_{k\alpha} \hat{c}_k
\end{eqnarray}
The current operators are thus given by 
\begin{equation}
\hat{I}_a = \sum_{k \in a} -i t_{k\alpha} \hat{d}^\dagger_k \hat{c}_\alpha + h.c.
\end{equation}

We solve for operators $\hat{c}_k$,

$$\hat{c}_k(t) = \hat{c}^0 e^{-i xi_k t} + \int d t' g_k(t,t') t_{k\alpha} \hat{d}_\alpha(t'),$$
$g_k(t,t') \equiv -i e^{-i\xi_k(t-t')}$, and subsequently for $\hat{d}_\alpha$,
$$
\hat{d}_\alpha(t) = \int dt' G_{\alpha\beta}(t,t') t^*_{\beta k} e^{-i \xi_k t'}\hat{d}_k^0
$$
where the Green's function obeys
\begin{equation}
\left( i \partial_t - \check{H} - \check{\Sigma}\right)\check{G} = \delta(t-t')
\end{equation}
and 
\begin{equation}
\check{\Sigma}(t,t') = \sum_k t^*_{k \alpha} g_k(t,t') t_{k \beta}.
\end{equation}
It is also useful to introduce partial $\Sigma$ that describe the decay to a certain lead,
\begin{equation}
\check{\Sigma}^a(t,t') = \sum_{k \in a} t^*_{k \alpha} g_k(t,t') t_{k \beta}
\end{equation}.
With this,
\begin{align}
\label{eq:c}
\hat{c}_k(t) = c^0_k e^{- i \xi_k t} \nonumber \\
+ g_k(t,t') t_{k \alpha} G_{\alpha \beta}(t',t'') t^*_{k'\beta} e^{-i\xi_{k'} t''} \hat{c}^0_{k'} 
\end{align}
in the above expression, we assume summation over $t',t'',k'$.
We substitute this into the current operator, average over the quantum state replacing $\langle \hat{c}^{0 \dagger}_k \hat{c}^0_k\rangle = f_k$  and get two contributions corresponding to two terms in Eq. \ref{eq:c}. 
The contribution $A$ depends only on the filling factor in the lead $a$ and reads 
\begin{equation}
I^a_A = {\rm Tr} \left( \check{G}(t,t') \check{F}^a(t',t) - \check{F}^a(t,t') \check{\bar{G}}(t,t') \right)
\end{equation}
where $\bar{G}(t,t') \equiv G^\dagger(t',t)$,
\begin{equation}
\check{F}^a(t,t') = \sum_{k \in a} t^*_{k\alpha}t_{k\beta} f_k e^{-i \xi_k(t-t')}
\end{equation}
The contribution B depends on filling factors in all leads
\begin{align}
I^a_B = {\rm Tr}\left( \check{G}(t,t') \sum_b \check{F}^b(t',t'') \check{\bar{G}}(t'',t''') \Sigma^\dagger_a(t''',t) \right. \nonumber \\
\left.- \Sigma_a(t,t') \check{G}(t',t'') \sum_b \check{F}^b(t'',t''') \check{\bar{G}}(t''',t) \right) 
\end{align}
We switch to the energy representation. To deal with the tunnel amplitudes, we will use the following relation
\begin{equation}
\label{eq:Gammadefinition}
\check{\Gamma}^a(\epsilon) = 2\pi \sum_k t^*_{k\alpha} t_{k \beta} \delta(\epsilon - \xi_k)
\end{equation}
$\check{\Gamma}^a$ characterizing the decay from all dots to the lead $a$. Conventionally, we will disregard the energy dependence of $\Gamma$ (since we are working close to the Fermi level). With this,
\begin{equation}
\check{F}^a = -i \check{\Gamma}^a f_a(\epsilon); \; \check{\Sigma}_a = - \frac{i}{2} \check{\Gamma}^a,
\end{equation}
where we have taken into account that the filling factor depends on energy only, and disregarded real part of $\Sigma$ (that would lead to a renormalization of the dot Hamiltonian). With this, the Green function is given by
\begin{equation}
\check{G} = \frac{1}{ \epsilon - \check{H} + i \check{\Gamma}/2};
\end{equation}
$\check{\Gamma} \equiv \sum_a \check{\Gamma}_a$.
The B contibution for the current for all $b \ne a$
can be written as
\begin{equation}
I_a/e = \sum_{b \ne a} \int \frac{d \epsilon}{2\pi} P_{a b} (\epsilon) f_b(\epsilon)
\end{equation}
$P_{ab}$ being the  probability to scatter from all channels of terminal $b$ to the channels of terminal $a$,
\begin{equation}
P_{ab}(\epsilon) = {\rm Tr} \{\check{\Gamma}^a \check{G}(\epsilon) \check{\Gamma}^b \check{\bar{G}}(\epsilon) \}
\end{equation}
This is in accordance with the corresponding part of Landauer formula for multi-terminal case. The contibution A reads:
\begin{equation}
I^a_A/e = -i \int \frac{d \epsilon}{2\pi} f_a(\epsilon) {\rm Tr} \{\check{\Gamma}^a (\check{G} -\check{\bar{G}})\}
\end{equation}
We use the relation
\begin{equation}
\label{eq:diffofgreens}
\check{G}  -\check{\bar{G}} = -i \check{G} \check{\Gamma} \check{\bar{G}}
\end{equation}
to represent the contribution A in the form
\begin{equation}
I^a_A/e = - \int \frac{d \epsilon}{2\pi} f_a(\epsilon) \sum_b P_{ab}(\epsilon)
\end{equation}
summing everything together, we reproduce the Landauer formula
\begin{equation}
I_a/e = \int \frac{d \epsilon}{2\pi} \sum_{b \ne a} P_{ab}(\epsilon)(f_a(\epsilon)-f_b(\epsilon))
\end{equation}
Let us construct a scattering matrix corresponding to the situation. The scattering to a terminal $a$ is described by $\check{\Gamma}^a$. Let us represent this matrix as $\check{\Gamma}^{a} = \check{W}_a^\dagger \check{W}_a$. The matrix $\check{W}_a$ is a matrix where the second index goes over the dots and the first one over the channels of the terminal $a$. This is of course an ambiguous representation, but so the scattering matrix is (tell more about?) We combine all matrices $W_a$ block by block to the matrix $W$ where the first index goes over all channels in all terminals. We note $\check{W}\dagger \check{W} = \check{\Gamma}$. With this, a scattering matrix describing the situation reads
\begin{equation}
\check{S} = 1 - i \check{W} \check{G} \check{W}^\dagger
\end{equation} 
Its unitarity can be proven with using the relation (\ref{eq:diffofgreens}).

\section{Normal transport for two dots}
\label{sec:twodots}
The case of the two dots, two terminals seems trivial but requires some elaboration for the limit where $ \Gamma$ in the dots are very different, this is the case under consideration. To warm up, let us specify to a single dot. We note that $\Gamma_a$ in this case are diagonal in spin owing to time-reversability and can be regarded as numbers. The transmission probability from the left to the right (or vice versa) 
can be written as 
\begin{equation}
T_0(\epsilon) = \frac{\Gamma_L \Gamma_R}{(\epsilon-E)^2 +\Gamma^2/4} 
\end{equation}
The ideal transmission is achieved at $\Gamma_L=\Gamma_R = \Gamma/2$ and $\epsilon = E$.
Let us go for two dots and list possible parameters of the model. Those are: level energies (split in spin) $E_1 + \boldsymbol{B}_1\cdot \boldsymbol{\sigma}$, $E_2 + \boldsymbol{B}_1\cdot \boldsymbol{\sigma}$, decays from the dots $\Gamma_1 = \Gamma^L_1 +\Gamma^R_1$, $\Gamma_2 = \Gamma^L_2 +\Gamma^R_2$, tunneling between the dots $\kappa + i \boldsymbol{\kappa} \cdot \boldsymbol{\sigma}$, and non-diagonal tunneling to the leads $\Gamma_{12,21} \equiv \gamma \pm i \boldsymbol{\gamma} \cdot \boldsymbol{\sigma}$. 
Let us write down the Green's function:
\begin{align}
\check{G}^{-1} = \epsilon 
- \begin{bmatrix}
H_1  &  H_{12} \\ 
H^\dagger_{12} & H_2
\end{bmatrix} \label{eq:twodotmodel}\\
H_{1,2} \equiv E_{1,2} + \boldsymbol{B}_{1,2}\cdot \boldsymbol{\sigma} -i\Gamma_{1,2}/2 \\
H_{12} \equiv \kappa + i \boldsymbol{\kappa} \cdot \boldsymbol{\sigma} -i(\gamma +i \boldsymbol{\gamma} \cdot \boldsymbol{\sigma})/2
\end{align}
The idea of further transform is that the second dot provides a featureless background for the first dot. To this end, we consider big $E_2,\Gamma_2 \gg \epsilon, B_2, E_1, \Gamma_1$
As to $\gamma,\kappa$, they are assumed to be of an intermediate scale, say $\gamma \simeq \sqrt{\Gamma_1 \Gamma_2}$. 

We will apply a transform that approximately diagonalises the Green function so that 
\begin{equation}
\check{G} = \check{U} \check{G}_d \check{U}^{-1}
\end{equation}
where 
\begin{align}
\check{U} =\sqrt{\frac{1+s}{2s}} \begin{bmatrix}1 & \eta_+\\
-\eta_- & 1\end{bmatrix};\; \\
\check{U}^{-1} =\sqrt{\frac{1+s}{2s}} \begin{bmatrix}1 & -\eta_+\\
\eta_- & 1\end{bmatrix}
\end{align}
and
\begin{align}
\eta_{\pm} &=& \frac{\mu_{\pm}}{1+s};\; s\equiv\sqrt{1+ \mu_+\mu_-}; \\
 \mu_{\pm} &=& 2 \frac{k \pm \boldsymbol{k}\cdot \boldsymbol{\sigma}  }{-E_2 + i\Gamma_2/2 }; \\
 k,\boldsymbol{k} &\equiv& -\kappa+i\gamma/2, - \boldsymbol{\kappa} + i \boldsymbol{\gamma}/2
\end{align}
with this, the biggest block of $\check{G}^{-1}_d$ is $-E_2 + i\Gamma_2/2$, while the smallest one reads
\begin{equation}
\epsilon - E_1 + i\Gamma_1/2 - \frac{k^2+\boldsymbol{k}^2}{-E_2 + i \Gamma_2/2}
\end{equation}
We rewrite it as
\begin{equation}
\epsilon - E_1 + i\Gamma/2 -\Delta E_1
\end{equation}
where the actual level width $\Gamma$ is given by 
\begin{align}
\Gamma &=& \Gamma_1 + \frac{\Gamma_2 C_{11} - 2 E_2 C_{10}}{E_2^2 + \Gamma_2^2/4}; \\
 C_{11} &\equiv& \kappa^2 - \gamma^2/4 + \boldsymbol{\kappa}^2 - \boldsymbol{\gamma}^2/4; \\ C_{10} &\equiv& \kappa\gamma + \boldsymbol{\kappa}\boldsymbol{\gamma}
\end{align} 
and we neglect insignificant shift of the level position 
\begin{equation}
\Delta E_1 = -\frac{C_{10}\Gamma_2/2 + C_{11} E_2}{E_2^2 + \Gamma_2^2/4}
\end{equation} 
The $\Gamma_a$ matrices are transformed as
$\check{\Gamma}^L \to \check{U}^\dagger \check{\Gamma}^L \check{U}$, $\check{\Gamma}^L \to \check{U}^{-1\dagger} \check{\Gamma}^L \check{U}^{-1}$. 

Keeing terms of the relevant orders only, we obtain
\begin{align}
\check{\Gamma}^L = \begin{bmatrix}
g_L& \Gamma_{12}^{+L} - \eta_-^* \Gamma_2^L \\
\Gamma_{12}^{-L} - \Gamma_2^L \eta_- & \Gamma_2^L 
\end{bmatrix}\\
g_L \equiv \Gamma_1^L - \Gamma_{12}^{+L} \eta_- -\eta_-^* \Gamma_{12}^{-L} + \eta_-^*\Gamma_2^L \eta_-
\end{align}

\begin{align}
\check{\Gamma}^R = \begin{bmatrix}
g_R & \Gamma_{12}^{+R} - \eta_+ \Gamma_2^R \\
\Gamma_{12}^{-R} - \Gamma_2^R \eta_+^* & \Gamma_2^R 
\end{bmatrix} \\
g_R \equiv \Gamma_1^R - \Gamma_{12}^{+R} \eta_+^* -\eta_+ \Gamma_{12}^{-R} + \eta_+\Gamma_2^R \eta_+^*- 
\end{align}

With this, we can  summarize the results for the total transmission coefficient $T_{tot}$ (summed over two spin directions). We introduce compact notations that adsorb the energy dependence of the coefficient:
\begin{align}
G_{\pm} = \frac{1}{\epsilon - E_1 \pm B +i\Gamma/2};\\ G_{s,a} = \frac{G_+\pm G_-}{2};\ \bar{G}_i = G^*_i
\end{align}
and write it down as
\begin{eqnarray}
T_{tot}(E)=2T_0 \nonumber \\
+(\Gamma_L \Gamma_R +\boldsymbol{\Gamma}^2) (G_+\bar{G}_+ + G_-\bar{G}_-) \\
+ 2((\boldsymbol{\Gamma}\cdot \boldsymbol{B})^2/B^2 -\boldsymbol{\Gamma}^2) G_a\bar{G}_a \\
+ RX (G_++G_-+\bar{G}_++\bar{G}_-) \\
- IX {\rm Im} (G_++G_--\bar{G}_+-\bar{G}_-) \label{eq:totalT}
\end{eqnarray}
Here, the partial decay rate read ($\Gamma_L+\Gamma_R =\Gamma$)
\begin{eqnarray}
\Gamma_L &=& \Gamma^L_1 +\frac{C_1 \Gamma^L_2-C^L_3 \Gamma_2 - 2 E_2 C_2^L}{E_2^2 +\Gamma_2^2/4} \\
C_1 &\equiv& \kappa^2 + \gamma^2/4 + \boldsymbol{\kappa}^2 + \boldsymbol{\gamma}^2/4  \\
C_2^L &\equiv& \boldsymbol{\kappa}\cdot \boldsymbol{\gamma}_L + \gamma_L \kappa \\
C_3^L &\equiv& \boldsymbol{\gamma}\cdot \boldsymbol{\gamma}_L +\gamma \gamma_L,
\end{eqnarray}
and similar for $R$.
The spin-orbit interaction is represented by the vector $\boldsymbol{\Gamma}$,
\begin{eqnarray}
\boldsymbol{\Gamma} &=& \frac{E_2 \boldsymbol{C}_5 +\boldsymbol{\kappa} C_4 + \boldsymbol{C}_6\times \boldsymbol{\kappa} +\kappa \boldsymbol{C}_6}{E_2^2 +\Gamma_2^2/4}\\
C_4 &= &\Gamma_2^L \gamma_R - \Gamma_2^R \gamma_L \\
\boldsymbol{C}_5 &=& \boldsymbol{\gamma}_R \gamma_L - \boldsymbol{\gamma}_L \gamma_R + \boldsymbol{\gamma}_R \times \boldsymbol{\gamma}_L \\
\boldsymbol{C}_6 &=& \Gamma_2^R \boldsymbol{\gamma}_L -\Gamma_2^L \boldsymbol{\gamma}_R
\end{eqnarray}
and the coefficients $RX$, $IY$ read
\begin{align}
RX &= \frac{1}{E_2^2 +\Gamma_2^2/4}(-E_2 C_7 +\kappa C_8 + \boldsymbol{\kappa} \cdot \boldsymbol{C}_9 &\nonumber\\
-&T_0 (E_2 C_{11} + C_{10} \Gamma_2/2))&\label{eq:RX}\\
IX &= \frac{1}{E_2^2 +\Gamma_2^2/4} (- C_7\Gamma_2/2 +\gamma C_8/2 + \nonumber &\\
&\boldsymbol{\gamma} \cdot \boldsymbol{C}_9/2 
 -T_0 (E_2 C_{10} - C_{11} \Gamma_2/2)) &\\
C_7 &= \gamma_R \gamma_L + \boldsymbol{\gamma}_R \cdot\boldsymbol{\gamma}_L
&\\
C_8 &= \Gamma_2^L \gamma_R + \Gamma_2^R \gamma_L\\
\boldsymbol{C}_9 &=  \boldsymbol{\gamma}_R \Gamma_2^L + \boldsymbol{\gamma}_L \Gamma_2^R &
\end{align}
We will explain the physical significance of each term in Eq. \ref{eq:totalT} in the next Section.

To treat the interaction self-consistently, we also need the average charge and spin in the dot,
\begin{equation}
\langle \hat{d}^\dagger_\alpha \hat{d}_\beta \rangle \equiv n \delta_{\alpha\beta} + \boldsymbol{n} \cdot \boldsymbol{\sigma} 
\end{equation}
This is given by 
\begin{align}
\check{n} = \int \frac{d\epsilon}{2\pi} \check{G}\left( (\Gamma_R + \boldsymbol{\Gamma}\cdot \boldsymbol{\sigma})f^R(\epsilon)\right. \nonumber\\
+\left.(\Gamma_L - \boldsymbol{\Gamma}\cdot \boldsymbol{\sigma})f^L(\epsilon) \right) 
\end{align} 
This can be rewritten in more detail as ($\boldsymbol{b} =\boldsymbol{B}/B$)
\begin{eqnarray}
n &=& \int \frac{d\epsilon}{2\pi} \left((G_s \bar{G}_s + G_a \bar{G}_a) (\Gamma_R f^R(\epsilon)  
 + \Gamma_L f^L(\epsilon))  \right.\nonumber \\
 &+& \left.  (\boldsymbol{b} \cdot \boldsymbol{\Gamma}) (G_a \bar{G}_s + G_s \bar{G}_a) (f^R(\epsilon)-f^L(\epsilon)) \right) \\
\boldsymbol{n} &=& \int \frac{d\epsilon}{2\pi} \left(2 \boldsymbol{b} (\boldsymbol{b}\cdot \boldsymbol{\Gamma}) G_a \bar{G}_a +\boldsymbol{\Gamma} (G_s \bar{G}_s - G_a \bar{G}_a) + \right. \nonumber\\ 
&& \left.
(\boldsymbol{b} \times \boldsymbol{\Gamma}) i (G_a \bar{G}_s - G_s \bar{G}_a)) (f^R(\epsilon)-f^L(\epsilon)) \right. \nonumber \\ &+& \left.
\boldsymbol{b} (G_a \bar{G}_s + G_s \bar{G}_a) (\Gamma_R f^R(\epsilon) + \Gamma_R f^R(\epsilon) \right)
\end{eqnarray}

We substitute filling factors at vanishing temperature $f^{L,R} = \Theta(eV_{L,R} - \epsilon)$ and integrate over $\epsilon$ to obtain $n, \boldsymbol{n}$ and full current. It is also advantageous at this stage to switch to dimensionless variables measuring energy in units of $\Gamma$ and setting $e=1$. We introduce convenient functions
\begin{eqnarray}
K^{\pm}_{R,L} &=& \frac{1}{2\pi} {\rm atan}(2(V_{R,L}-\epsilon_d \pm B)); \\
L^{\pm}_{R,L} &=& \frac{1}{2\pi} {\rm ln}( 4(V_{R,L} \pm B)^2+1);\\
L^{\pm} &=& L^{\pm}_{R} - L^{\pm}_{L}; \; K^{\pm} = K^{\pm}_{R} - K^{\pm}_{L}.
\end{eqnarray}
With this, 
\begin{eqnarray}
n&=& \sum_{k=L,R} \Gamma_k (1/2 +K^+_k +K^-_k)  \nonumber \\
&+&(\boldsymbol{b} \cdot \boldsymbol{\Gamma}) (K^++K^-), \\
\boldsymbol{n} &=& \boldsymbol{b} \left( \Gamma_R (K^-_R -K^+_R) + \Gamma_L (K^-_L -K^+_L)\right) \nonumber\\
+& \frac{\boldsymbol{\Gamma} }{1+4 B^2}& \left(K^++K^- +B(L^--L^+)\right) \nonumber\\
+& \frac{(\boldsymbol{b} \times \boldsymbol{\Gamma}) }{2(1+4 B^2)} & \left(B(K^++K^-) +L^+-L^-\right) \nonumber \\
+& \frac{2 \boldsymbol{b} (\boldsymbol{b} \cdot \boldsymbol{\Gamma}) B}{1+4 B^2}&\left(4B(K^++K^-) +L^+-L^-\right) 
\end{eqnarray}
The self-consistency equations then read:
\begin{equation}
\epsilon_d = U n; \; \boldsymbol{B} = \boldsymbol{B}_0 - U \boldsymbol{n}
\end{equation}
$\boldsymbol{B}_0$ being the external magnetic field. This equation has to be solved at each $V_{R,L}$. With this solution, we can evaluate the current 
\begin{eqnarray}
I &=&  T_0(V_L-V_L)/\pi + 2 (\Gamma_L\Gamma_R +\boldsymbol{\Gamma}^2) (K^++K^-) \nonumber \\
&+& 4 ((\boldsymbol{\Gamma}\cdot \boldsymbol{b})^2 - \boldsymbol{\Gamma}^2)\frac{B}{1+4 B^2} \times \nonumber \\
& \times & \left(4B(K^++K^-) +L^+-L^-\right) \nonumber \\
&+& RX (L^++L^-) - IX (K^++K^-)/2 \label{eq:currentself}
\end{eqnarray} 

\begin{figure}[tb]
\begin{center}
	\includegraphics[width=0.4\textwidth]{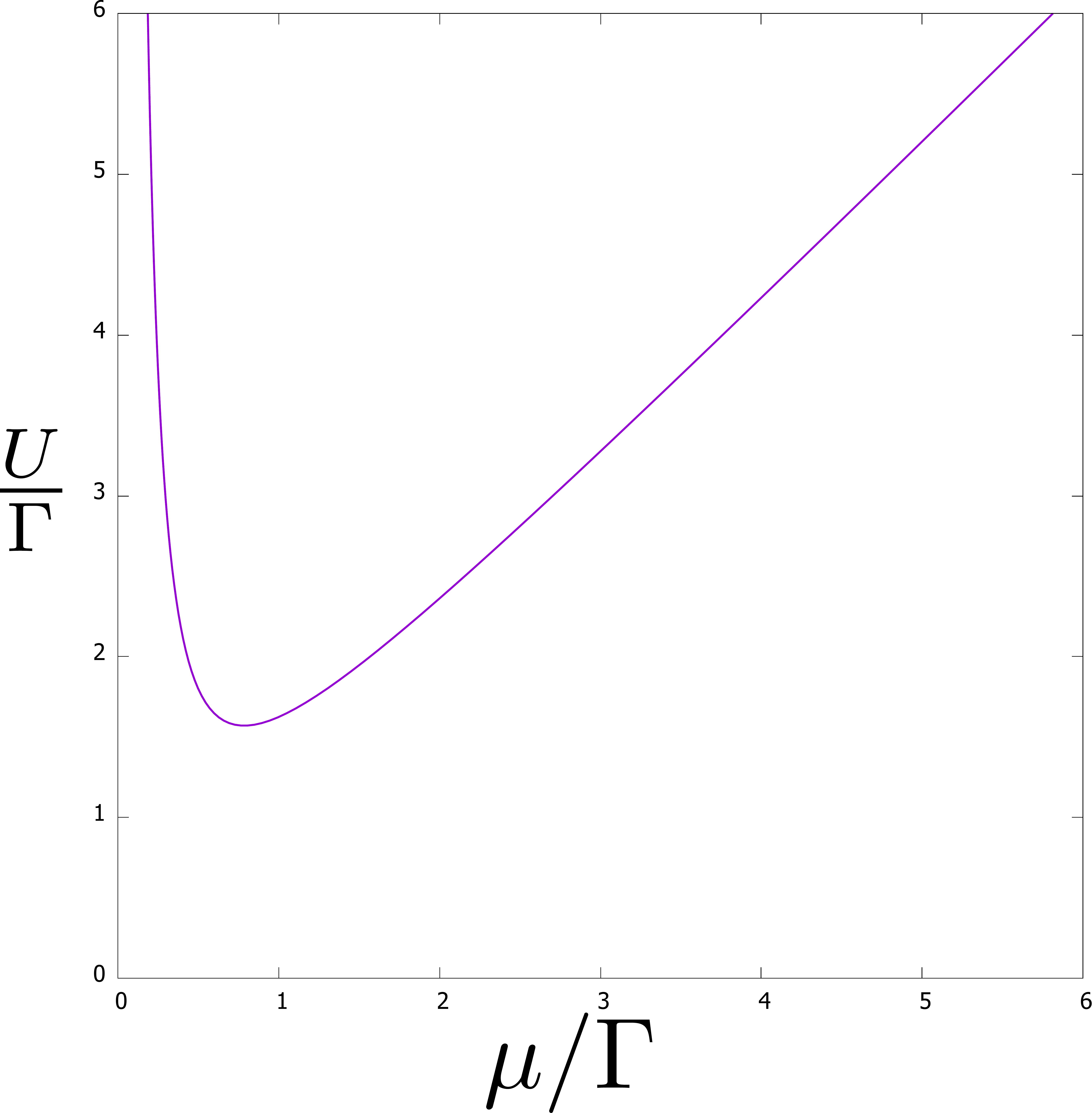}
\end{center}
\caption{The domain of magnetic phase.}
\label{uversusmuplot}
\end{figure}

Let us elaborate on the equilibrium case $V_R=V_L=\mu$. The terms with spin-orbit interaction do not appear in this case and the self-consistency equations read ($\tilde{K}=K^R=K^L$)
\begin{align}
\epsilon_d = U (1/2 + \tilde{K}^+ +\tilde{K}^-); \\
 \boldsymbol{B} =\boldsymbol{B}_0 -\boldsymbol{b} U (\tilde{K}_- - \tilde{K}_+)
\end{align}
We specify to $\boldsymbol{B}_0 =0$ and determine the boundary of spontaneously magnetic phase where $B \to 0$. In this limit,
\begin{equation}
\tilde{K}_- - \tilde{K}_+ \to  - B\frac{2}{\pi} \frac{1}{1+4(\mu^*)^2}; \mu^* = \mu - \epsilon_d
\end{equation}
with this, the equations for the boundary read
\begin{align}
U= (1+4(\mu^*)^2)\frac{\pi}{2}; \\
 \mu = \mu^* + U (1/2 + (1/\pi) {\rm atan}(2 \mu^*))
\end{align}
An implicit plot is given in Fig \ref{uversusmuplot}. The splitting occurs above critical value $U_c = \pi/2$, at large $U$ the magnetic phase occurs in the interval $\mu = (0,U)$ as it should be.

\section{ Normal transport examples}
\label{sec:normaltransportexamples}
In this Section, we will analyse the peculiarities of normal transport in the model at hand. We restrict ourselves to zero-voltage conductance and non-interacting case where zero-voltage conductance is simply given by $T_{tot}$ at $\epsilon$ corresponding to Fermi level,
\begin{equation}
G(V_g) = \frac{G_Q}{2} T_{tot}(\epsilon=E_F).
\end{equation}
Since $E_1$ is a linear function of the gate voltage, and shift of $\epsilon$ in Eq. \ref{eq:totalT} is equivalent to the shift of $E_1$, the energy dependence of $T_{tot}$ directly gives the gate voltage dependence of the conductance. The conductance with interaction is qualitatively similar to the non-interacting one since the main effect of interaction in our model is the spin-splitting corresponding to $B \simeq U$.

Let us explain the physical significance of the terms in Eq. \ref{eq:totalT}. All spin-orbit effects are incorporated into a single vector $\boldsymbol{\Gamma}$ in the spin space. To start with, let us neglect the spin-orbit interaction setting $\boldsymbol{\Gamma}=0$, so we can disregard the third term. In this case, $T_{tot}$ is contributed independently by spin orientations $\pm$ with respect to $\boldsymbol{B}$. Their contributions are shifted by $2B$ in energy.

The first term in Eq. \ref{eq:totalT} gives the featureless transmission of the transport channel and asymptotic value of the conductance at $|E_1| \gg \Gamma$. The second term describes the resonant transmission via the localized state and would show up even if there is no interference between the transmissions through the channel and the localized state. It rives rise to a Lorentzian peak - resonant transmission - of the width $\simeq \Gamma$ in conductance that splits into two at sufficiently big spin splitting $\simeq \Gamma$. Let us bring the fifth term into consideration. Since $G-\bar{G} = -i \Gamma G\bar{G}$ its energy dependence is identical to the second one. However, it usually gives a negative contribution to transmission describing destructive interference of the transmissions in the dot and in the channel - resonant reflection. 

The fourth term describes the celebrated Fano effect coming about the interference of the resonant and featureless transmission. It is visually manifested as asymmetry of otherwise Lorentzian peaks or dips. The antisymmetric Fano tail $\propto \epsilon^{-1}$ at large distances from the peak/dip centre beats Lorentzian tail $\propto \epsilon^{-2}$. All these terms are hardly affected by spin-orbit interaction, while the second one manifests it fully. It mixes up spin channels and makes conductance to depend on the orientation of $\boldsymbol{B}$ with respect to $\boldsymbol{\Gamma}$. 

\begin{figure*}
\begin{center}
	\includegraphics[width=\textwidth]{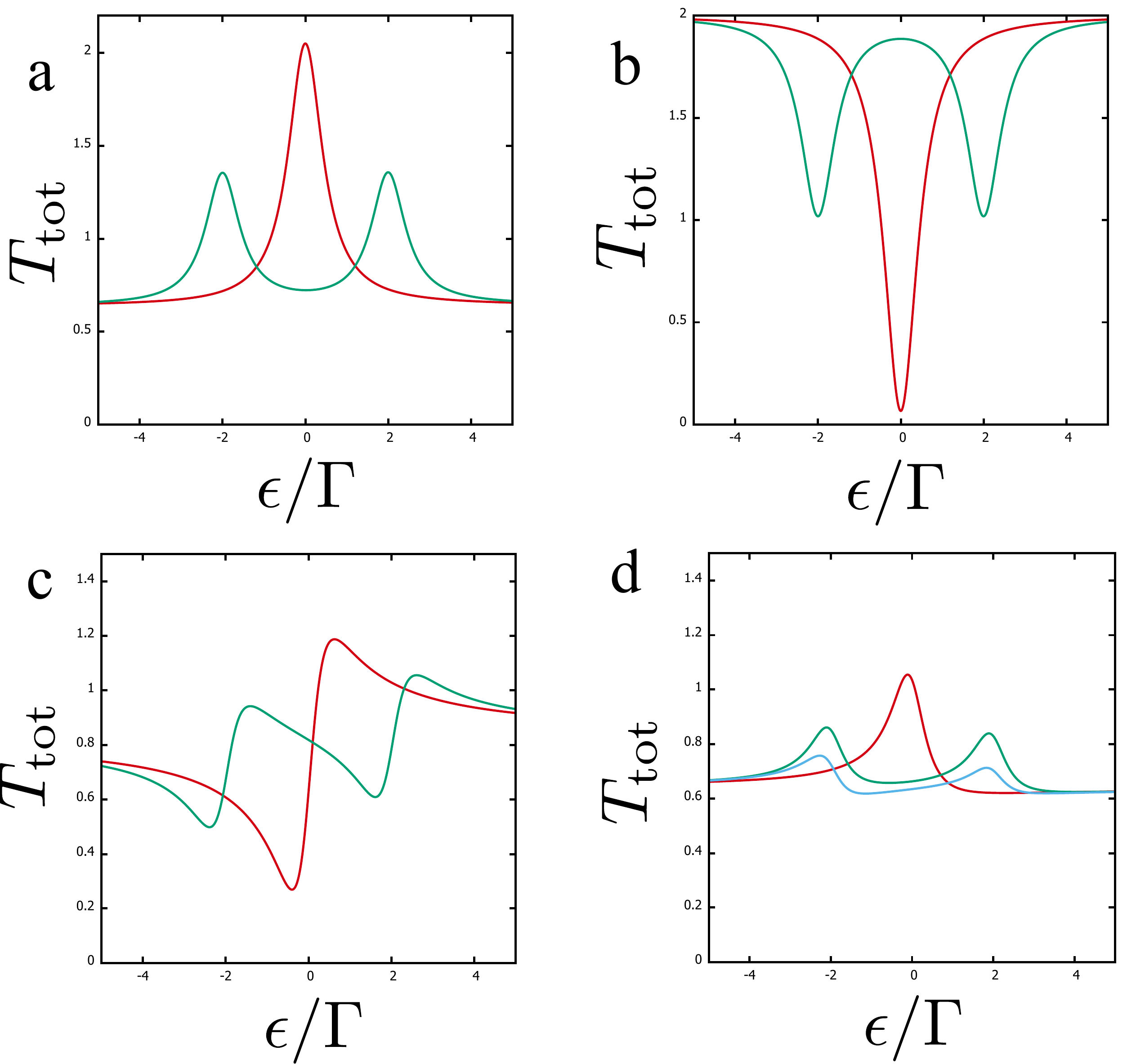}
\end{center}
\caption{Examples of normal transport. The energy dependence of $T_{tot}$ is the same as the conductance dependence on the gate voltage. Red curves correspond to $B=0$, green curves to $B=2\Gamma$.  a. Basic example: resonant transmission b. Dip: resonant reflection c. Fano. d. Strong spin-orbit. Here, green (blue) curve is for parallel (perpendicular) orientation of $\boldsymbol{B}$ with respect to $\boldsymbol{\Gamma}$. }
\label{fig:normaltransport}
\end{figure*}

We illustrate the possible forms of the conductance energy/gate-voltage dependence with the plots in Fig. \ref{fig:normaltransport} for 4 settings of the parameters $\Gamma_2^{L,R}, E_2, \kappa, \boldsymbol{\kappa}, \gamma_{L,R}, \boldsymbol{\gamma}_{L,R}$. Owing to separation of the scales assumed, the relevant parameters $\Gamma_{L,R}, \boldsymbol{\Gamma}, RX, IX$ are invariant with respect to rescale with the factor $A$,
\begin{align}
\Gamma_2^{L,R}, E_2 \to A (\Gamma_2^{L,R}, E_2) \\
\kappa, \boldsymbol{\kappa}, \gamma_{L,R}, \boldsymbol{\gamma}_{L,R} \to \sqrt{A} (\kappa, \boldsymbol{\kappa}, \gamma_{L,R}, \boldsymbol{\gamma}_{L,R}).
\end{align}
For all settings, energy is in units of the resulting $\Gamma$. For each setting, we give the plots at $B=0$ and $B = 2 \Gamma$, the latter to achieve a visible separation of resonant peculiarities. Spin-orbit interaction is weak except the last setting where we give separate plots for $\boldsymbol{B} \parallel \boldsymbol{\Gamma}$ and $\boldsymbol{B} \perp \boldsymbol{\Gamma}$. 

For Fig. \ref{fig:normaltransport} a we choose $\Gamma_2^L,\Gamma_2^R,E_2 = A(0.2,0.8,0.5)$, $\kappa, \gamma_L,\gamma_R = \sqrt{A}(0.5,0.2,0.2)$, $\Gamma_1^L, \Gamma_1^R = 1.6, 3.5$. We also specify small but finite spin-orbit terms yet they hardly affect the conductance. In this case, the transmission through the localized state is faster than the interference with the transmission in the channel. This results in a resonant reflection peak at $B=0$ that splits into two upon increasing the magnetic field. A little Fano asymmetry can be noticed upon a close look.

For Fig. \ref{fig:normaltransport} b we choose $\Gamma_2^L,\Gamma_2^R,E_2 = A(0.5,0.5,0)$, $\kappa, \gamma_L,\gamma_R = \sqrt{A}(3.5,0.2,0.2)$, $\Gamma_1^L, \Gamma_1^R = 0.5, 0.5$.
The transmission trough the channel is ideal, $T_0 =1$, the localized state is connected to the channel better than to the leads. This results in a pronounced resonant reflection dip at $B=0$ that also splits into two upon increasing the magnetic field.

For Fig. \ref{fig:normaltransport} c we choose $\Gamma_2^L,\Gamma_2^R,E_2 = A(0.2,1.5,0)$, $\kappa, \gamma_L,\gamma_R = \sqrt{A}(1.5,0.3,0.1)$, $\Gamma_1^L, \Gamma_1^R = 0.8, 0.1$. This choice is such that the competing processes of resontant transmission and reflection almost compensate each other so the resulting resonance peculiarity assumes almost antisymmetric Fano shape. The separation of the peculiarities upon the spin splitting is less pronounced than in the previous examples owing to long-range Fano tails mentioned.

We illustrate the effect of strong spin-orbit interaction in Fig. \ref{fig:normaltransport} d.  We choose $\Gamma_2^L,\Gamma_2^R,E_2 = A(0.2,0.8,0.5)$, $\kappa, \gamma_L,\gamma_R = \sqrt{A}(0.5,0.2,0.2)$, $\Gamma_1^L, \Gamma_1^R = 1.6, 3.5$. 
As to spin-dependent parameters, we choose 
$\boldsymbol{\kappa} = \sqrt{A} SO [0, 0.2, -0.6]$, $\boldsymbol{\gamma}_L = \sqrt{A} SO [0.3,0,0]$, $\boldsymbol{\gamma}_R = \sqrt{A} SO [0.0,0,1]$ and set the coefficient $SO$ to $1.6$, this is its maximal value that satisfies the positivity conditions imposed on the matrices of the rates. The peculiarity at $B=0$ is a peak with a noticeable Fano addition. It splits at $B= 2 \Gamma$ changing its shape, that is different for $\boldsymbol{B} \parallel \boldsymbol{\Gamma}$ and 
$\boldsymbol{B} \perp \boldsymbol{\Gamma}$ as well as for positive and negative energies. Note that owing to Onsager symmetry $
G(\boldsymbol{B}) = G(-\boldsymbol{B})$.

\begin{figure*}
\begin{center}
	\includegraphics[width=\textwidth]{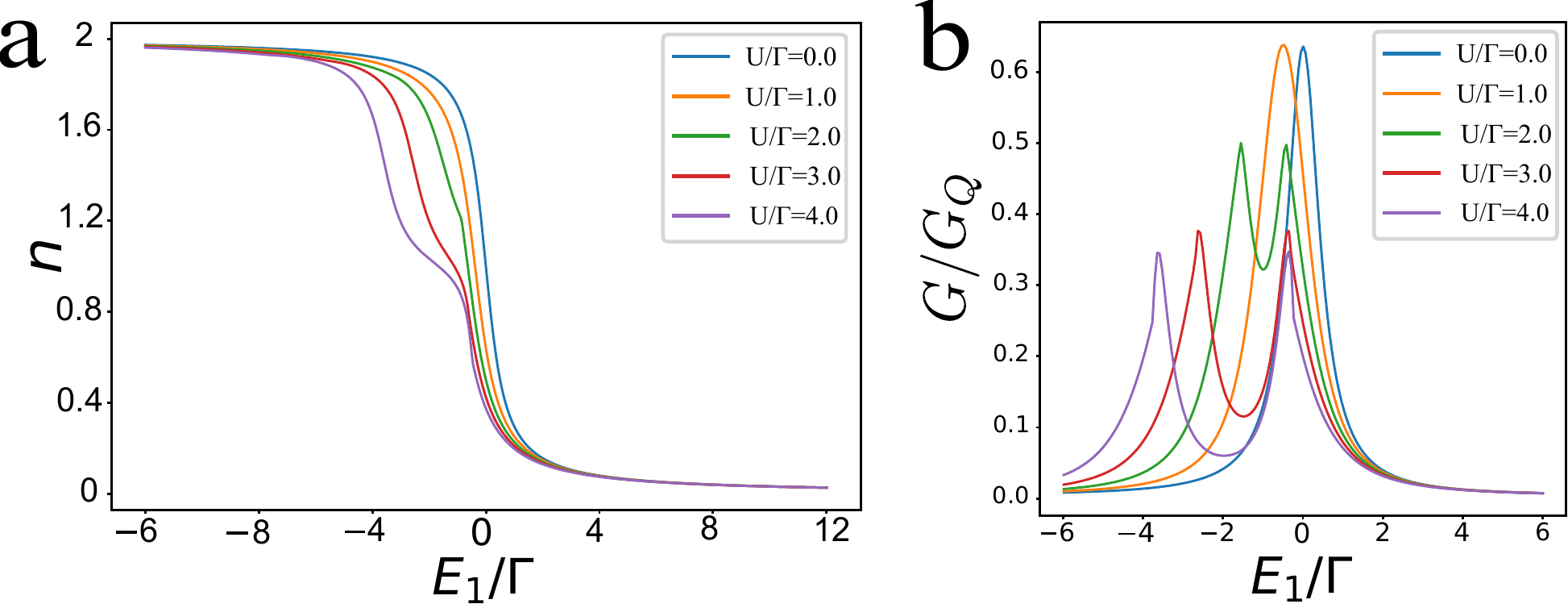}
\end{center}
\caption{Example of normal transport with interaction. Resonant transmission regime, no magnetic field, no SO coupling. The setup parameters are given in the text.  a. The average number of electrons in the localized state versus $E_1$ at various interaction strengths. b. Zero-voltage conductance versus $E_1$ at various interaction strengths. }
\label{fig:self}
\end{figure*}

We also provide an example with interaction implementing the self-consistent scheme described in the previous Section (Fig. \ref{fig:self}). For this example, we choose $\Gamma_2^L,\Gamma_2^R,E_2 = A(0.2,1.5,-15)$, $\kappa, \gamma_L,\gamma_R = \sqrt{A}(0.8,0.1,0.1)$, $\Gamma_1^L, \Gamma_1^R = 1.1, 0.9$. This choice corresponds to very low channel transmission ($T_0 = 10^{-3}$).
The average number of electrons in the dot is presented in Fig. \ref{fig:self}a as a function of $E_1$ for several interaction strengths, at zero voltage difference and magnetic field. All curves change from full occupation at big negative $E_1$ to zero occupation big positive $E_1$. At $U=0$ and $U=\Gamma$ the curves are smooth with no 
spontaneous spin splitting emerging throught the whole interval of $E_1$. For higher interaction strengths, there is an interval of $E_1$ where the spontaneous splitting is present. The ends of this interval are in principle manifested by cusps in the curves. Only cusps at the end of the interval close to zero are visible, the cusps at the other end are too small.
It might seem that the zero-voltage conductance (Fig. \ref{fig:self} b) can be computed from $T_{tot}$ at the parameters  $\tilde{E}_1, \tilde{B}$ that solve the self-consistency equation at zero voltage difference. However, this is not so, since these parameters also depend on voltage difference. We compute  zero-voltage conductance by numerically differentiating the current (Eq. \ref{eq:currentself}) at small voltage differences. At zero interaction, we see a resonant transmission peak. Its height does not reach $G_Q$ because of the asymmetry $\Gamma_R \ne \Gamma_L$. At $U=\Gamma$, there is still a single peak. At higher $U$ we see the splitting of the peak. The height of the peaks split is a half of the height of the original peak if they are sufficiently separated. As we have conjectured earlier, this is qualitatively similar to the conductance traces where spin splitting is induced by the magnetic field. 

\section{Superconducting transport}
\label{sec:suptrans}
In this Section, we elaborate on the description of superconducting transport in our model. Since supercurrent is a property of the ground state of the system, it is convenient to work with electron Green functions in imaginary time and introduce Nambu structure. Let us start, as we did previously, with an arbitrary number of dots and superconducting leads. If we neglect tunnel couplings, the inverse Green function ${\cal H}(\epsilon)$ is a matrix in the space of the dots, spin and Nabmu and reads:
\begin{equation}
\check{{\cal H}} = i \epsilon \tau_z - \check{H}.
\end{equation}
The tunnel couplings to the leads labelled by $a$ add the self-energy part
\begin{align}
\check{{\cal H}} = i \epsilon \tau_z - \check{H} + \frac{i}{2} \sum_a \check{\Gamma}_a \check{Q}_a
\end{align}
where $\check{\Gamma}_a$ are given by Eq. \ref{eq:Gammadefinition} and the matrix $\check{Q}_a$ is a matrix in Nambu space reflecting the properties of the superconducting lead $a$,
\begin{equation}
Q_{a}=\frac{1}{\sqrt{\epsilon^2 +\Delta_a^2}}
\begin{bmatrix}
\epsilon& \Delta_a e^{i\phi_{a}}\\
\Delta_a e^{-i\phi_{a}}& -\epsilon
\end{bmatrix},
\end{equation}
$Q_a^2=1$.

To find supercurrents, we need to evaluate the total energy
and take its derivatives with respect to the phase differences. Since that are dots that connect the leads with different phase, the phase-dependent energy is the energy of the dots. The latter can be expressed as

\begin{equation}
	{\cal E} =-\frac{1}{2}\int \frac{d\epsilon}{2 \pi} \ln \det(\check{{\cal H}})
\end{equation}

To see how does this work, let us check this formula neglecting tunnel couplings.
With this, the energy is the sum over eigenvalues of $\check{H}$, $E_n$,
\begin{equation}
	{\cal E} =-\frac{1}{2}\int \frac{d\epsilon}{2 \pi} \ln (\epsilon^2 + E_n^2)
\end{equation}
The integral formally diverges at $\epsilon \to \infty$. To regularize it, we 
substract its value at $E_n=0$ to obtain
\begin{equation}
{\cal E} = - \sum_n \frac{|E_n|}{2} + const
\end{equation}
To recover a familiar formula, we shift the constant by ${\rm Tr}(\check{H})/2$,
 \begin{align}
{\cal E} = - \sum_n \frac{|E_n|}{2} + \sum_n \frac{E_n}{2} + const =\\
\sum_n E_n \Theta(-E_n) + const, 
\end{align}
so it becomes the energy of the filled states (those with $E_n<0)$.
This suggest we need to handle the integral with care keeping eye on possible problems at big $\epsilon$.
Fortunately, no special care has to be taken for the phase-dependent energy since it is accumulated at superconducting gap scale $\epsilon \simeq \Delta$. We have to be careful when expressing the occupation of the dots in terms of derivatives of ${\cal E}$ with respect to dot energies (as we do for numerical calculations). For instance, the average occupation of the dot $1$ reads
\begin{equation}
\langle \hat{n}_1 \rangle = \frac{\partial {\cal E}}{\partial E_1} + 1,
\end{equation} 
the last term correcting for high-energy divergences.

For our starting two-dot, two-lead model, the inverse Green function reads (c.f. with Eq. \ref{eq:twodotmodel}). 

\begin{equation}
{\cal H}=\begin{bmatrix}
{\cal H}_{11}&{\cal H}_{12}\\
{\cal H}_{21}&{\cal H}_{22}
\end{bmatrix}
\end{equation}
, where 

\begin{align}
{\cal H}_{11}&=i\epsilon \tau_z-E_1  - (\boldsymbol{B}_1 \cdot \check{\boldsymbol{\sigma}})\tau_z \nonumber\\
+ & \frac{i}{2} (\Gamma_{1}^{R} \check{Q}_R + \Gamma_{1}^{L} \check{Q}_L),\label{eq:H11}\\
{\cal H}_{22}&=i\epsilon \tau_z-E_2  - (\boldsymbol{B}_1 \cdot \check{\boldsymbol{\sigma}})\tau_z  \nonumber \\ 
+  &\frac{i}{2} (\Gamma_{2}^R \check{Q}_R + \Gamma_{2}^L \check{Q}_L),\\
{\cal H}_{12}&=-\check{\kappa} +\frac{i}{2} \{ \check{\gamma}_{L} \check{Q}_L+\check{\gamma}_R \check{Q}_R \},\\
{\cal H}_{21}&=-\check{\kappa}^{\dagger}+\frac{i}{2} \{ \check{\gamma}_{L}^\dagger \check{Q}_L+\check{\gamma}_R^\dagger \check{Q}_R \},
\end{align}

and we turn back to the compact notations

\begin{align}
\check{\kappa},\check{\kappa}^\dagger&=\kappa\pm i \boldsymbol{\kappa}\cdot\boldsymbol{\sigma}\\
\check{\gamma}_{L,R},\check{\gamma}_{L,R}^\dagger&=\gamma_{L,R}\pm i \boldsymbol{\gamma_{L,R}}\cdot\boldsymbol{\sigma}
\end{align}
and made use of $Q$ matrices corresponding to two leads
\begin{equation}
\check{Q}_{L,R}=\frac{1}{\sqrt{\epsilon^2 +\Delta^2}}
\begin{bmatrix}
\epsilon& \Delta e^{i\phi_{L,R}}\\
\Delta e^{-i\phi_{L,R}}& -\epsilon
\end{bmatrix}.
\end{equation}


Next goal is to reduce the number of parameters implementing the separation of scales mentioned and implemented for the normal transport.  This is achieved by the following transformation of the determinant
\begin{align}
\ln \det(\check{{\cal H}})=\ln \det(\check{{\cal H}}_{11}-\check{{\cal H}}_{12}\check{{\cal H}}_{22}^{-1}\check{{\cal H}}_{21}) \nonumber\\ + \ln \det(\check{{\cal H}}_{22})
\end{align}
and implementing $E_2,\Gamma_2 \gg \gamma,\kappa \gg \epsilon, B_2, E_1, \Gamma_1$.

Let us first evaluate $\det(\check{{\cal H}}_{22})$, which is that of a $4\times 4$ matrix with spin structure taken into account. Since we may assume $\epsilon, B_2 \ll \Gamma_2, E_2$ 
the spin structure is trivial and the answer reads 


\begin{align}
\ln \det(\check{{\cal H}}_{22})=2 \ln ( E_2^2 +\frac{1}{4} \Gamma_2^2 )+ \nonumber\\ 2 \ln(1-T_0 \frac{\Delta^2}{\Delta^2 +\epsilon^2 } \sin^2{\phi/2}),
\end{align}
where, as previously, we define $\Gamma_2=\Gamma_{2}^{L}+\Gamma_{2}^{R}$ and $T_0=\Gamma_{2}^{L}\Gamma_{2}^{R}/( E_2^2 +\frac{1}{4} \Gamma_2^2 )$.

The energies of Andreev levels are determined from zeros of this determinant. We recover the well-known expression for the energy of the spin-degenerate Andreev level in a contact with transparency $T_0$,
\begin{equation}
E_{Andr} = \Delta \sqrt{1 - T_0 \sin^2(\phi^/2)}
\end{equation}
The integration of the log of the determinant over the energy gives the expected result for the energy of the ground state,
\begin{equation}
{\cal E} = - E_{Andr}
\end{equation}

Let us turn to evaluation of the rest of the expression. We note that 
\begin{equation}
\check{{\cal H}}^{-1}_{22} = -\frac{E_2  + \frac{i}{2} (\Gamma_{2R} \check{Q}_R + \Gamma_{2L} \check{Q}_L)}{\Big( E_2^2+\frac{\Gamma_2^2}{4}\Big)(1-T_0 s)} 
\end{equation}
where we have introduced a convenient compact notation 
\begin{equation}
s \equiv \frac{\Delta}{\sqrt{\Delta^2 +\epsilon^2}} \sin^2(\phi^/2)
\end{equation}
The matrix in the first determinant thus contains a factor $(1-T_0 s)$ in the denominator. Multiplying with this factor cancels $\det(\check{\cal H}_22)$ so the whole expression can be reduced to the following relatively simple form
\begin{align}
\ln \det(\check{{\cal H}})= \label{eq:finalsup}\\
\ln \det \left( 
(1 -T_0s)(i\epsilon \tau_z-E_1  - (\boldsymbol{B} \cdot \check{\boldsymbol{\sigma}})\tau_z) \right.\nonumber \\
+\Delta E + s \Delta E_S \\
+ \frac{i}{2} (\Gamma^R(s) \check{Q}_R + \Gamma^{L}(s) \check{Q}_L)  \\
+\left.\frac{i}{4} \boldsymbol{\Gamma}\cdot \check{\boldsymbol{\sigma}} (\check{Q}_L \check{Q}_R - \check{Q}_R \check{Q}_L)\right),\nonumber
\end{align}
where
\begin{align}
\Gamma^{L,R}(s) = \Gamma^{L,R} + s \Gamma_S^{L,R}.
\end{align}
The parameters $\Gamma_{L,R}$, $\Delta E$, $\boldsymbol{\Gamma}$ have been already defined in our consideration of normal transport. The compact description of superconducting transport brings three additional parameters
\begin{align}
\Delta E_S = \frac{ - E_2 C_7 + \kappa C_8 + \boldsymbol{\kappa} \cdot \boldsymbol{C}_9}{E_2^2 + \Gamma_2^2/4} \\
\Gamma_S^{L}= - T_0 \Gamma_1^L + \frac{\Gamma_R (\gamma_L^2 + \boldsymbol{\gamma}_L^2)}{E_2^2 + \Gamma_2^2/4} \\
\Gamma_S^{R}= - T_0 \Gamma_1^R + \frac{\Gamma_L (\gamma_R^2 + \boldsymbol{\gamma}_R^2)}{E_2^2 + \Gamma_2^2/4}.
\end{align}
Here, $\Delta E_S$ is a part of the expression (\ref{eq:RX}) for $RX$ but is an independent parameter. 

Since both normal and superconducting transport originate from the same scattering matrix, there are many examples when the parameters characterizing the superconducting transport can be directly determined from the results of normal transport measurements, a single channel with transparency $T_0$ being the simplest one. The presence of the additional parameters $\Delta E_S$, $\Gamma_S^{L,R}$is therefore rather disappointing: we cannot predict superconducting transport exclusively from the results of normal transport measurements and have to rely on model assumptions.

Let us outline the physical meaning of the overall structure of the expression (\ref{eq:finalsup}). The first term is a product of the terms whose zeros give the Andreev level in the transport channel and energy level in an isolated localized state, the product indicate that these levels are independent. The rest of the terms thus describe the hybridization of these levels. Note that the terms with $\Delta E$ cannot be cancelled by a shift of $E_1$, so in distinction from the normal case, are active in the presence of superconductivity. The terms with $\Gamma(s)$ are similar to tunnel decay terms in Eq. \ref{eq:H11}, in distinction from normal case the presence of the second dot does not just renormalize $\Gamma$. The last term describes spin-orbit effect and is proportional to the same vector $\boldsymbol{\Gamma}$ as in the normal case. In distinction from all other terms, it is odd in the phase difference since it is proportional to the commutator of two $\check{Q}$. The combination of this term and that with magnetic field results in a shift of the mimimum of superconducting current from $0$ or $\pi$ positions.


\section{Numerical details}
\label{sec:numerics}
In this Section, we provide the overal strategy and details of our numerical calculations.

We postpone the discussion of self-consistency assuming that we already know $E_1$ and $\boldsymbol{B}$. To find the phase-dependent energy, we have to integrate the log of the determinant over $\epsilon$. We compute directly the determinant of $8 \times 8$ matrices implementing the difference of scales numerically.  
For quick computation at each energy, we represent the matrix $\check{{\cal H}}$ as a sum over various scalar functions of $\epsilon$,

\begin{equation}
\check{{\cal H}} =\check{A}+\epsilon \check{B} +\frac{\epsilon}{\sqrt{\epsilon^2+\Delta^2}} \check{C} + \frac{\Delta}{\sqrt{\epsilon^2+\Delta^2}} \check{D}(\phi_L,\phi_R)
\end{equation}
where the matrices $\check{A} - \check{D}$ do not depend on $\epsilon$ and only $\check{D}$ depends on the supercondictig phase.
 We define the function of $\epsilon$ as 
 $\log(\det(\check{{\cal H}}(\epsilon,\phi=0))-\log(\det(\check{{\cal H}}(\epsilon,\phi=0))$
 and intregrate using scipy.quad. Direct evaluation of the sum over discrete equidistant $\epsilon$ the interval of the order of $5 \Delta $ provides comparable numerical efficiency and accuracy. 

As mentioned, we treat interaction self-consistently, as the interaction-induced shift in $E_1$ and $\boldsymbol{B}$. The self-consistency equations read as

\begin{align}
&\tilde{E_1}=E_1+U n(\tilde{E_1},\tilde{\boldsymbol{B}})\\
&\tilde{\boldsymbol{B}}=\boldsymbol{B}-U \boldsymbol{n}(\tilde{E_1},\tilde{\boldsymbol{B}})
\end{align}

,where the average number of particles on the dot are given by derivatives of the total energy
$N=(\partial_{E_1} {\cal E} +1) $ and $\boldsymbol{n}=\partial_{\boldsymbol{B_1}}{\cal E}$.
We compute these derivatives integrating the analytical derivatives of $\det(\check{{\cal H}}(\epsilon)$ over $\epsilon$. These integrals may converge at a $\epsilon \gg \Delta$ provided $E_{1},\boldsymbol{B} \gg \Delta$, an adaptive grid of discrete $\epsilon$ could be chosen to speed up the evaluation, yet using scipy.quad suffices for our purposes.

To solve the self-consistency equations, we implement a root-finding  minimization algorithm minimizing the function  $F=f^2+|\boldsymbol{f}|^2$, where $f, \boldsymbol{f}$ are defined as 
\begin{align}
&f=\tilde{E_1}-E_1-U n(\tilde{E_1},\tilde{\boldsymbol{B}})\\
&\boldsymbol{f}=\boldsymbol{\tilde{B}}-\boldsymbol{B}+U \boldsymbol{n}(\tilde{E_1},\tilde{\boldsymbol{B}})
\end{align}
and checking if the minimum is achieved at $F=0$. 
Alternatively, we can make use of the fact that the solutions of the self-consistency equations correspond to the extrema of the following energy functional 
\begin{equation}
E_{T}(\tilde{E}_1,\tilde{\boldsymbol{B}})= {\cal E}(\tilde{E}_1,\tilde{\boldsymbol{B}}) -\frac{(E_1 - \tilde{E}_1)^2}{2 U} + \frac{(\boldsymbol{B}- \tilde{\boldsymbol{B}})^2}{2 U}
\end{equation}
Unfortunately, this energy functional is not bounded, and the extrema required are rather saddle points than minima. However, they can be found, for instance, by maximization of the functional in $\tilde{E}_1$ and subsequent minimization in $\tilde{\boldsymbol{B}}$.


The Andreev bound states are given by the zeros of the  determinant at imaginary $\epsilon$ (thus real energy $E = i\epsilon$)
in the interval $(0,\Delta)$. We find these roots by minimizing $\det(\check{{\cal H}}(\epsilon))^2$ and checking if the minimum is achieved at zero. but because of the different scales in the problem and existence of a big scale, we first try to find an equivalent matrix, whose determinant is more numerically apt to minimize. 
Typically, there are multiple Andreev states, so we subdivide the interval $(0,\Delta)$ to find them all.

\section{Superconducting transport examples}
\label{sec:supexamples}

In this Section, we present the examples of numerical evaluation of supercurrent and Andreev state energies in our model, for 3 sets of parameters.
It has to be noted that the parameter space of the model is large and at the moment the examples are not exhaustive. We have been mostly searching for the parameter sets manifesting a pair of $0-\pi$ transitions.
This is achieved if $E_T(\phi=\pi) - E_T(\phi=0)\equiv E_\pi <0$. Contrary to our initial expectations, it is rather difficult to achieve such inversion for an arbitrary parameter set at high transmission $T_0$. It is rather easy to find $0-\pi$ transitions at low $T_0$. More extensive exploration of the parameter space is required to reveal the role of various parameters and draw conclusions. Nevertheless, the following examples provide interesting illustrations of rich physics captured by the model.

In this Section and in all plots, we measure the energies,decay rates and the current $I/2e$ in units of $\Delta$. 

\begin{figure*}[hb]
	\centering
	\begin{subfigure}{0.45\textwidth}
		\includegraphics[width=\textwidth]
{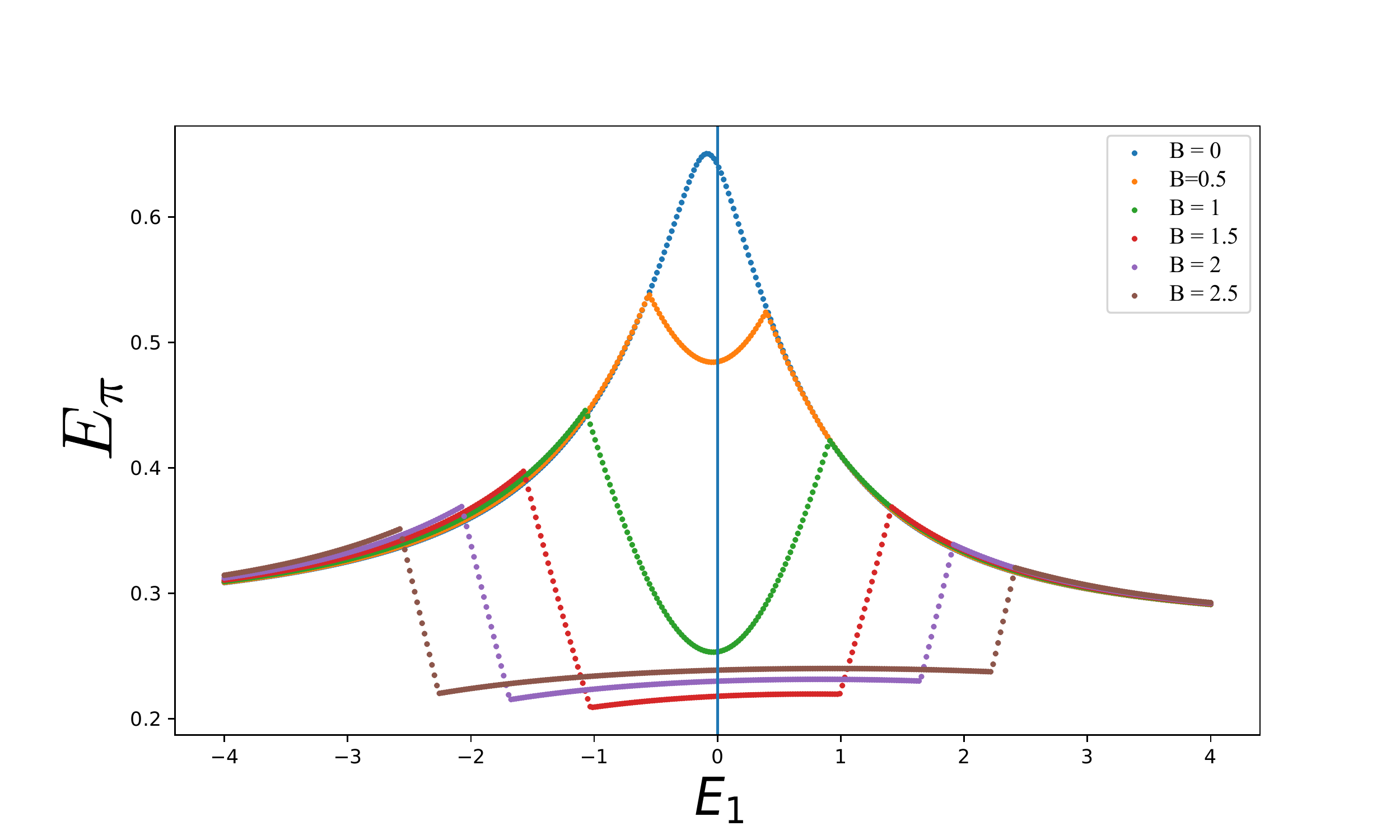}
			\caption{$E_\pi \equiv E_T(\phi=\pi) - E_T(\phi=0)$ versus $E_1$ at several values of magnetic field.}
			\label{fig:E_T_symmetric}
		\end{subfigure}
		\begin{subfigure}{0.45\textwidth}
			\includegraphics[width=\textwidth]
{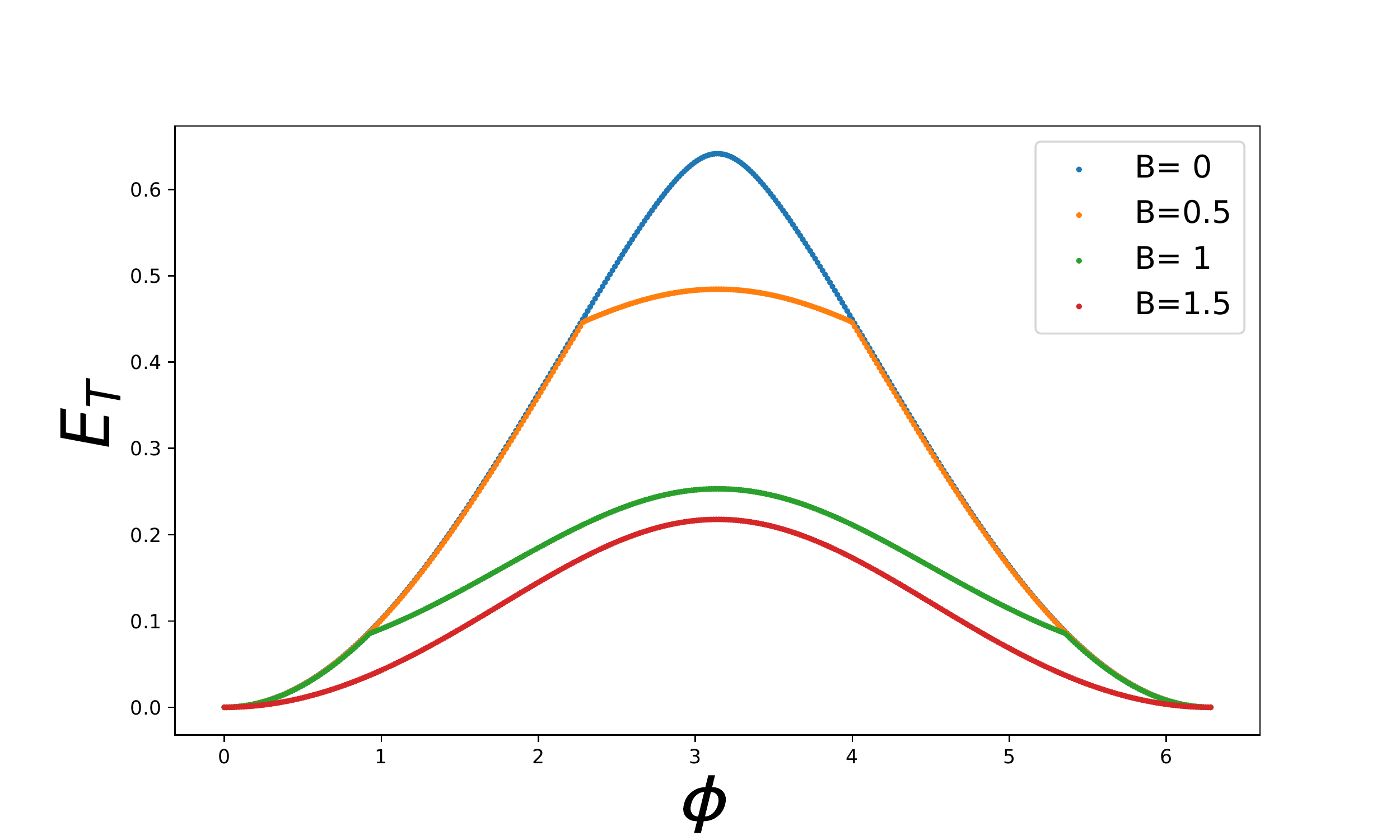}
			\caption{The phase-dependent part of energy $E_T \equiv E_T(\phi) - E_T(\phi=0)$ at $E_1=0$ and several values of magnetic field.}
			\label{fig:E_T_symmetric1}
		\end{subfigure}
		\begin{subfigure}{0.45\textwidth}
			\includegraphics[width=\textwidth ]
{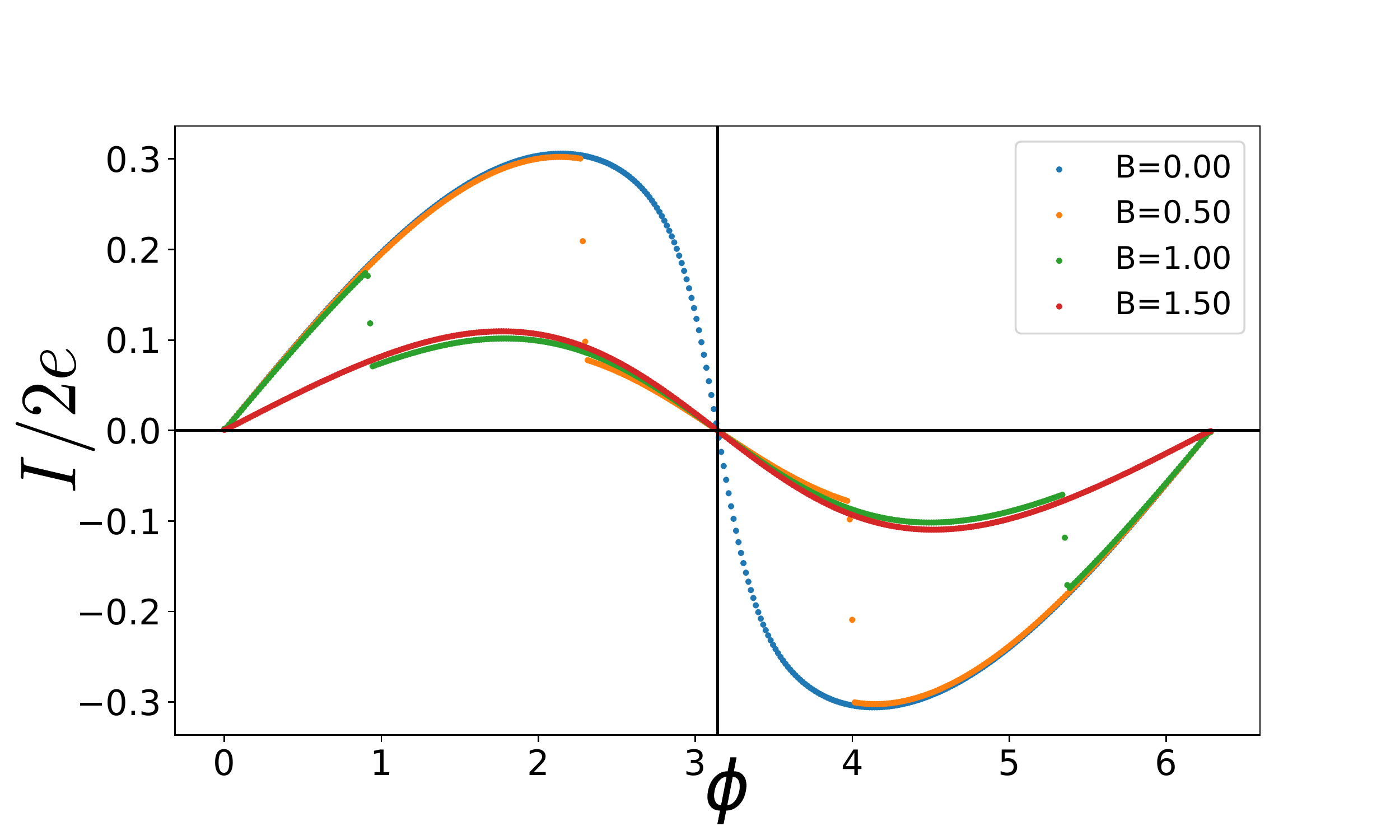}
			\caption{The phase dependence of the superconducting current at $E_1=0$ for several values of $B$.}
			\label{fig:I_symmetric}
		\end{subfigure}
		\begin{subfigure}{0.45\textwidth}
			\includegraphics[width=\textwidth]
{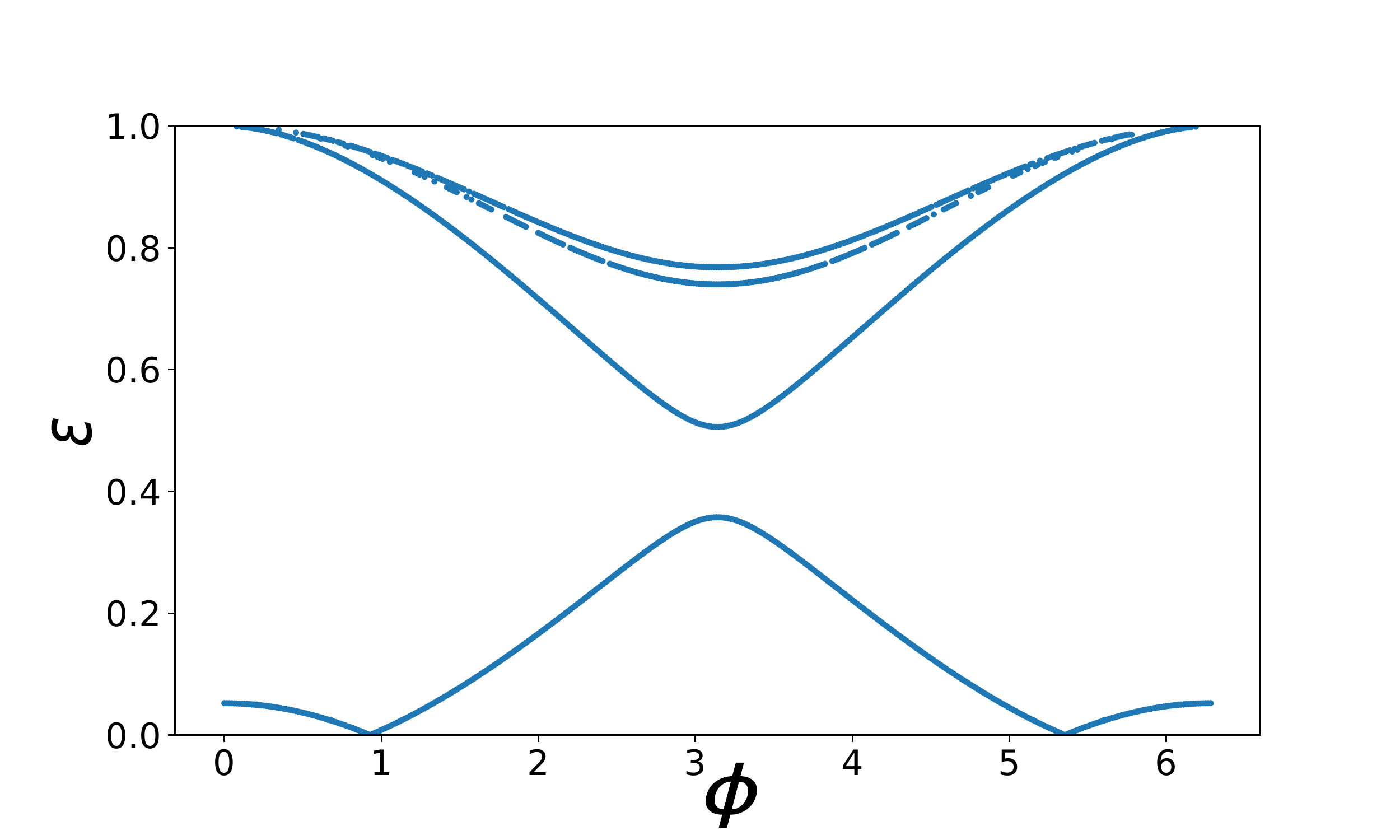}
			\caption{The phase dependence of ABS energies at $E_1=0$ and $|B|=1$.}
			\label{fig:ABS_symmetric}
		\end{subfigure}
	\begin{subfigure}{0.45\textwidth}
		\includegraphics[width=\textwidth]
{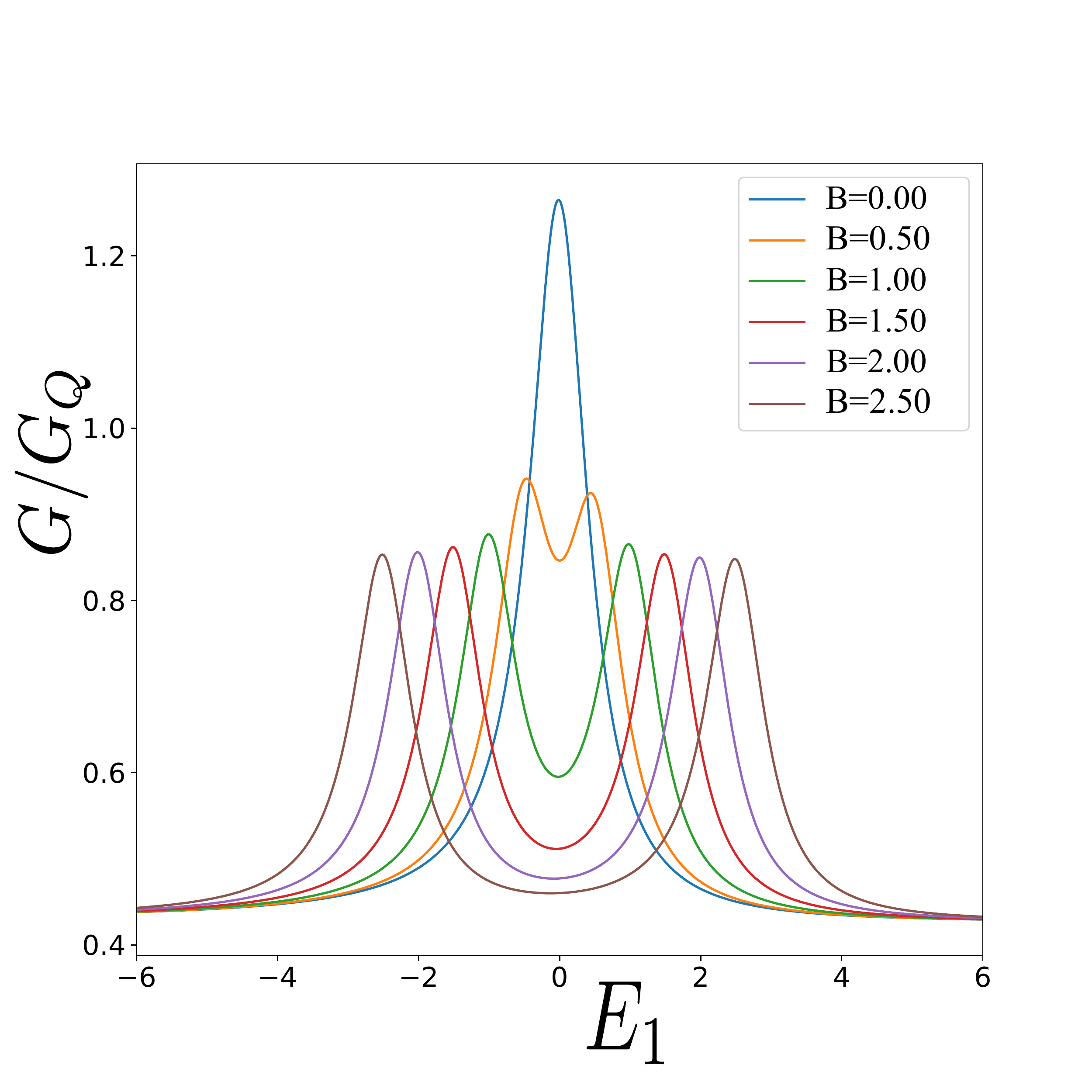}
		\caption{Normal zero-voltage conductance versus $E_1$ at several values of magnetic field.}
		\label{fig:ABS_symmetric2}
	\end{subfigure}
	\caption{Example A. Resonant transmission, moderate channel transmission. No SO coupling. }
	\label{fig:Hristo0}
	\end{figure*}

{\it Example A} (Fig. \ref{fig:Hristo0})
Here, we disregard spin-orbit coupling and interaction. The parameters are $\Gamma_2^L,\Gamma_2^R,E_2 = A(1.9,1.9,-2)$, $\kappa, \gamma_L,\gamma_R = \sqrt{A}(0.4,0.1,0.1)$, $\Gamma_1^L, \Gamma_1^R = 1.2, 0.9$
and correspond to $T_0 = 0.47$, $\Gamma_L =1.3 , \Gamma_R =0.96, \Gamma =2.22$.
As we can see from the normal conductance traces presented in Fig. \ref{fig:ABS_symmetric2}, for this parameter set we have the resonant transmission accompanied by very weak Fano asymmetry. The resonant transmission peak splits upon increasing the magnetic field. 

Actually, this set illustrates an unsuccessful attempt to achieve a pair of $0=\pi$ transitions. This is seen from the plots in Fig. \ref{fig:E_T_symmetric} that give the (gate-voltage) traces of $E_\pi$ at various magnetic fields. Zero-field trace peaks near the centre of conductance peak indicating the enhancement of supercurrent upon increasing the transmission, and saturates at finite value at $|E_1| \gg 1$: this saturation is achieved for all magnetic fields. Upon increasing the magnetic field the value of $E_T$ near the resonance gets down. It becomes smaller than the saturated value at $B > 0.8$. It seems it has a chance to pass zero manifesting $0-\pi$ transitions. However, this does not happen:the tendency changes and the minimum of $E_T$ starts to increase at $B > 1.5$. Prominent features in the plots are sharp cusps in energy dependence. They indicate the crossings of Andreev states with zero energy that, for the features in this plot, are located at $\phi=\pi$ and corresponding gate voltages.

Let us set $E_1=0$ and look at the phase dependence of energy(Fig \ref{fig:E_T_symmetric1}) for a set of magnetic fields. Here we also see the cusps corresponding to the crossings at certain values of the phase. Superconducting currents plotted in Fig.\ref{fig:I_symmetric} are obtained by numerical differentiation of the energy, so the cusps become jumps, the discontinuities of the current. The zero-field curve is prominently non-sinusoidal as expected for high transmission at this value of $E_1$. The current becomes smaller tending to almost sinusoidal curve at high magnetic fields upon increasing magnetic field, but does not get inverted. At intermediate fields, the current jumps between non-sinusoidal and sinusoidal curve.

In Fig. \ref{fig:ABS_symmetric} we show the phase dependence of ABS energies for $E_1=0$ and $|B_1|=1$. We see four spin-split ABS counting from down up. Eventually, the picture of ABS demonstrates little interference between the transport channel and the localized state. The third and the fourth curves are close to $E_{Andr}$ for $T_0=0.47$ and are thus associated with the transport channel, their spin-splitting $\simeq 0.1$ is small coming from the interference. The first and the second state are associated with the dot. The spin splitting is thus big: the first curve  looks like the second curve shifted by $\simeq 1$ downwards, with the part shifted to negative energy being flipped to positive ones. This also explains sharp cusps in the first curve.

Although in our model the phase-dependent energy is not a minus half-sum of ABS energies as it would be for energy-independent transmission, we can use this sum for qualitative estimations. With this, the half-sum of the first and second energies would result in an inverted supercurrent, but the half-sum of the third and fourth states, that is, the contribution of the transport channel, adds to the balance a usual supercurrent of slightly bigger amplitude.

\begin{figure*}
	\centering
	\begin{subfigure}{0.45\textwidth}
		\includegraphics[width=\textwidth]
{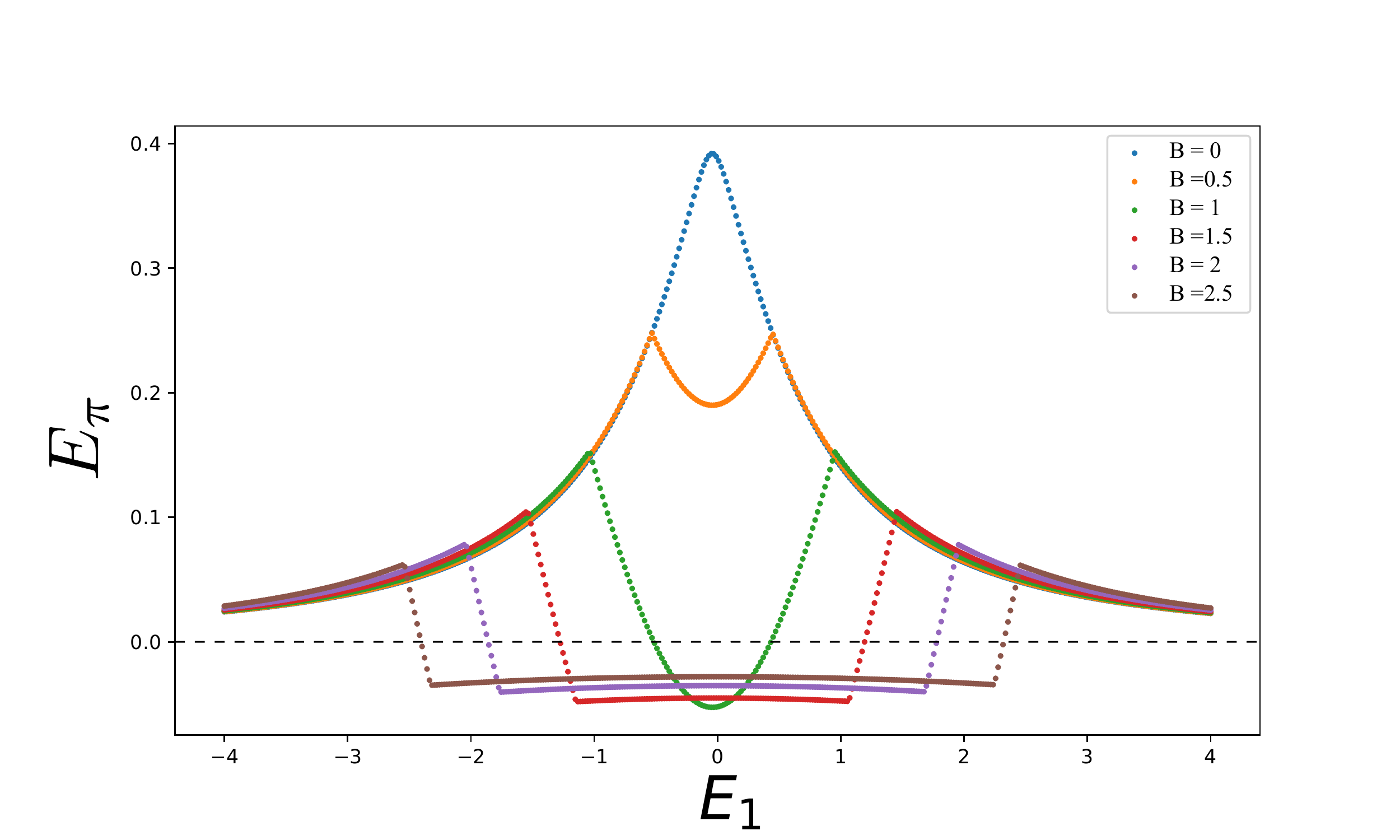}
		\caption{$E_\pi$ versus $E_1$ at several values of magnetic field.}
		\label{fig:E_T_inversion}
	\end{subfigure}
	\begin{subfigure}{0.45\textwidth}
		\includegraphics[width=\textwidth ]
	{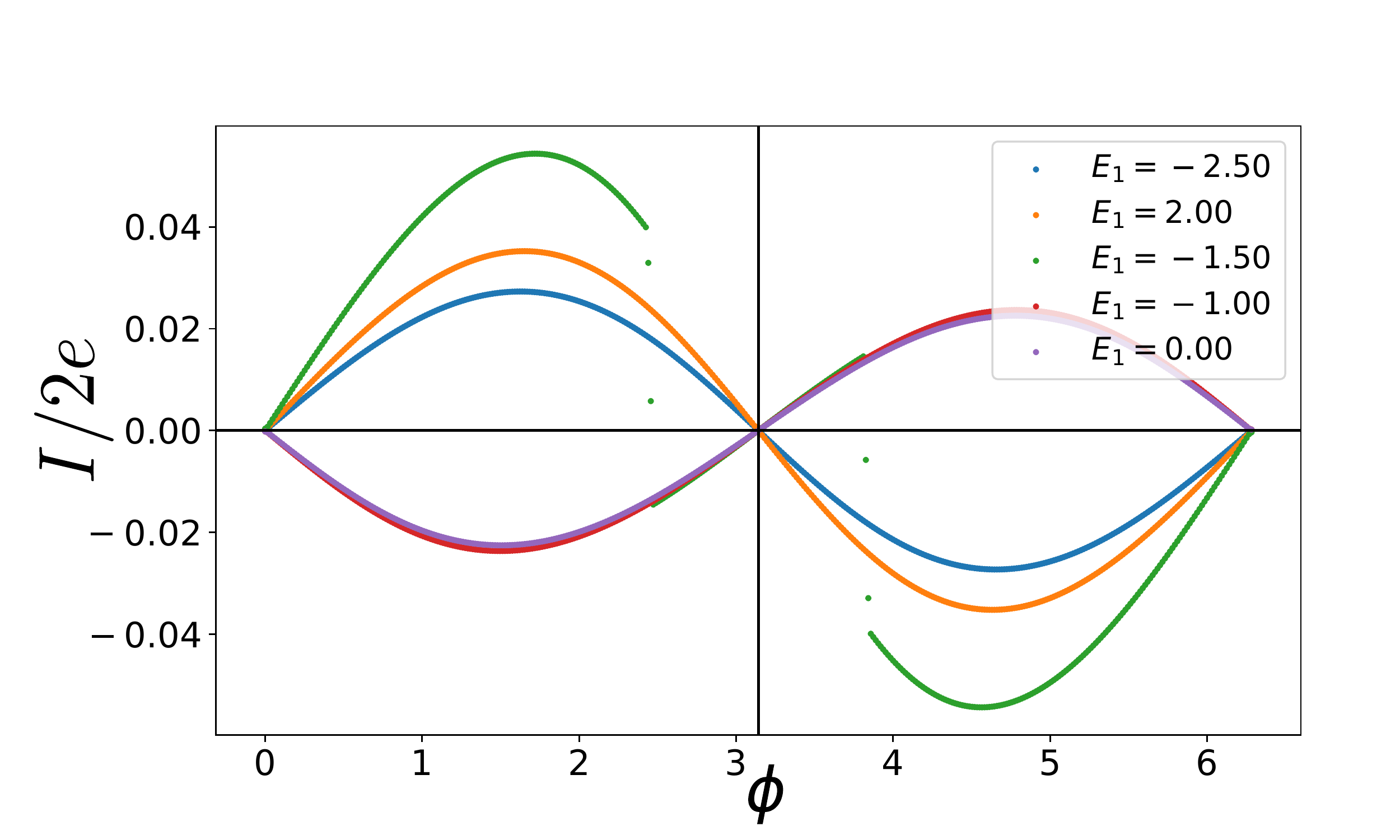}	
		\caption{The phase dependence of the superconducting current at $|B|=1.5$ for several values of $E_1$.}
		\label{fig:I_inversion}
	\end{subfigure}
	\begin{subfigure}{0.45\textwidth}
		\includegraphics[width=\textwidth]
		{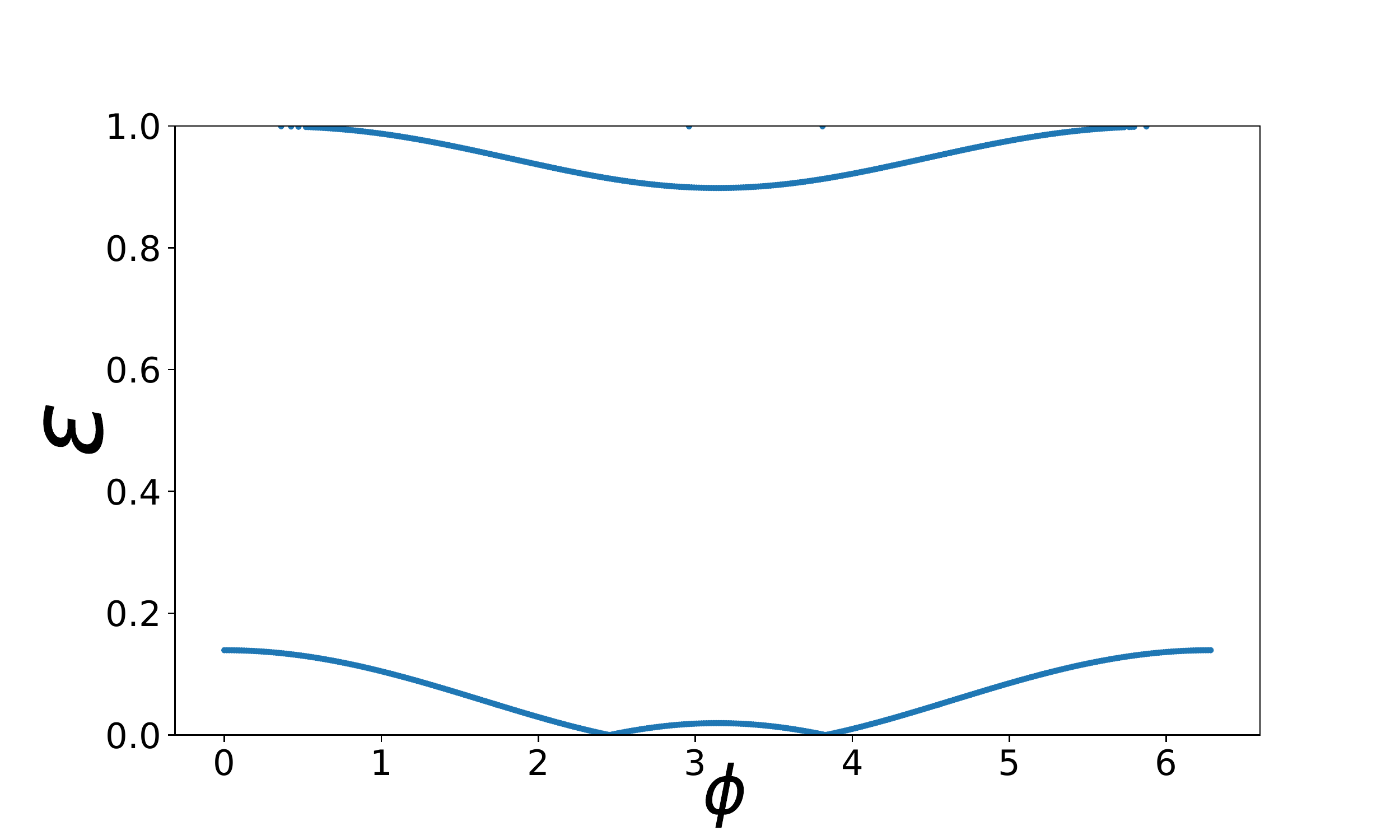}
		\caption{The phase dependence of ABS energies at $E_1=-1.5$ and $|B|=1.5$.}
		\label{fig:ABS_inversion}
	\end{subfigure}
	\begin{subfigure}{0.45\textwidth}
	\includegraphics[width=\textwidth]
	{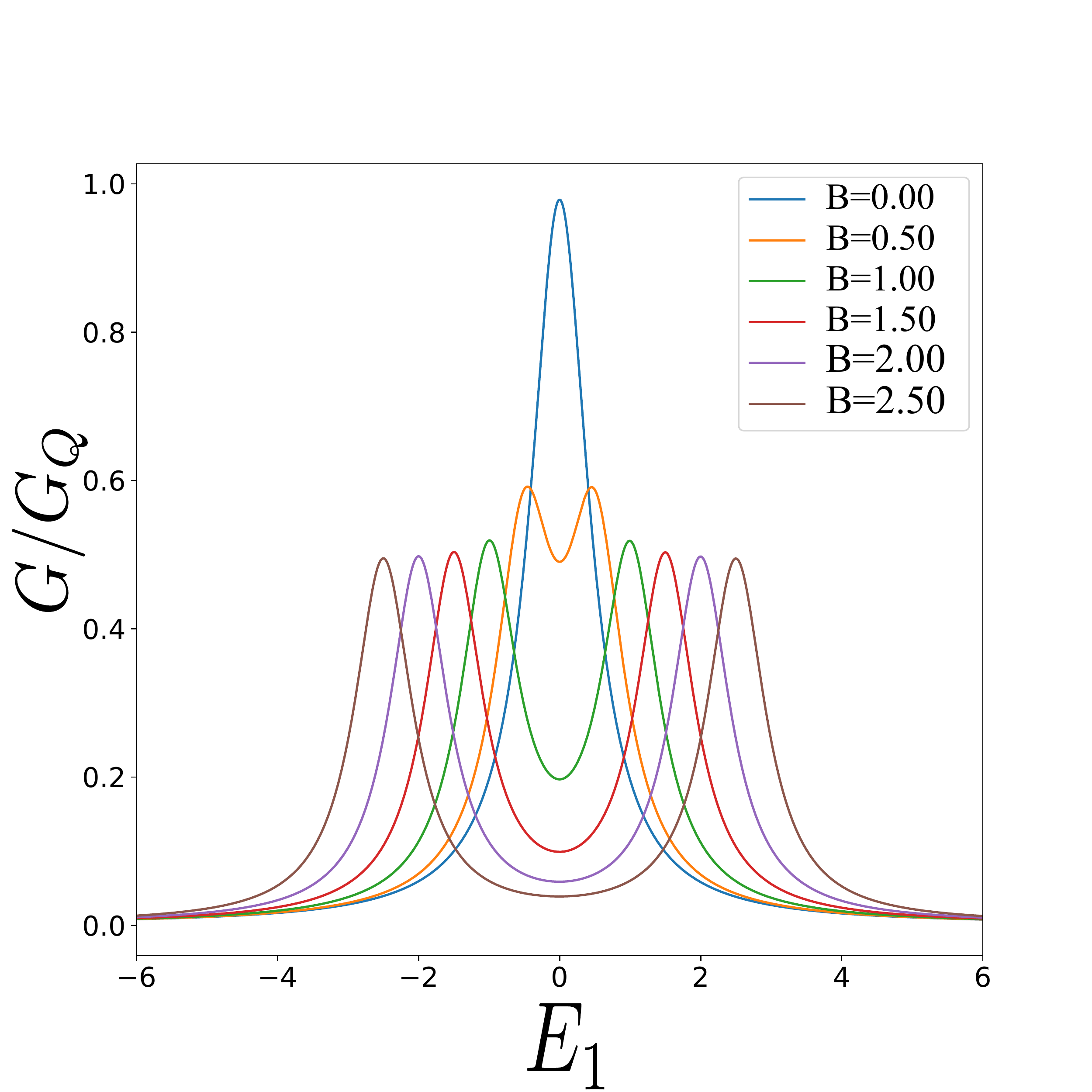}
	\caption{Normal zero-voltage conductance versus $E_1$ at several values of magnetic field.}
	\label{fig:Normal_conductance_symmetry}
\end{subfigure}
\caption{Example B. Resonant transmission, low channel transmission. No SO coupling. A pair of $0-\pi$ transitions occurs at $|B|>0$.}
\label{fig:Hristo1}
\end{figure*}

{\it Example B}. (Fig. \ref{fig:Hristo1}) This inspired us to check if the $0-\pi$ transitions can be achieved at very low transmission of the transport channel. We have taken the following set of parameters $\Gamma_2^L,\Gamma_2^R,E_2 = A(0.2,1.5,-15)$, $\kappa, \gamma_L,\gamma_R = \sqrt{A}(0.8,0.1,0.1)$, $\Gamma_1^L, \Gamma_1^R = 1.1, 0.91$. For this choice, $T_0\simeq 10^{-3}$  $\Gamma_L=1.1, \Gamma_R = 0.91, \Gamma=2.03$. 
The normal conductance traces (Fig. \ref{fig:Normal_conductance_symmetry}) 
show a classical scenario of resonant transmission that saturates to almost zero far from the resonance. 
 
The check was successful.
We plot the traces of  $ E_T \equiv E_T(\phi=\pi) - E_T(\phi=0)$ for various magnetic fields in Fig. \ref{fig:E_T_inversion}. The traces look like those in Fig. \ref{fig:E_T_symmetric} except the shift downwards by $\simeq 0.25$. Owing to this, $E_T$ is negative for $B> 0.8$ in an interval of gate voltages that increases with $B$, $0-\pi$ transitions are at the ends of the interval. 

We plot the phase dependence of the supercurrent for $|B|=2$ and various $E_1$ in Fig. \ref{fig:I_inversion}. The $0-\pi$ transitions at this field take place at $E_1 \approx \pm 1.25$. In accordance with this, the almost sinusoidal curves at $E_1 = -2.5, 2$ are of positive amplitude while those at $E_1 = 0, -1$ are of negative one. Note a rather low value $\simeq 0.02$ of the maximum "negative" current, almost 25 times smaller than the maximum value of the current in a single transport channel. An interesting curve is found close to the transition, at $E_1 = -1.5$. Here, the current jumps between sin-like curves of positive and negative amplitude. The total integral of the current between $0$ and $\pi$ is still positive, so $E_\pi >0$. 

An example of the phase dependence of ABS energies is given in Fig. \ref{fig:ABS_inversion}. Since the transmission of the channel is very low, we see only two spin-split ABS. The upper one is close to the gap edge, and eventually merges with continuous spectrum at $\phi \approx 0.6, 2\pi - 0.6$. The lower one is close to zero, and exhibits two zero crossings at $\phi \approx \pi \pm 0.65$ corresponding to the discontinuities in corresponding curve in Fig. \ref{fig:I_inversion}.

\begin{figure*}
	\centering
	\begin{subfigure}{0.45\textwidth}
		\includegraphics[width=\textwidth]
		{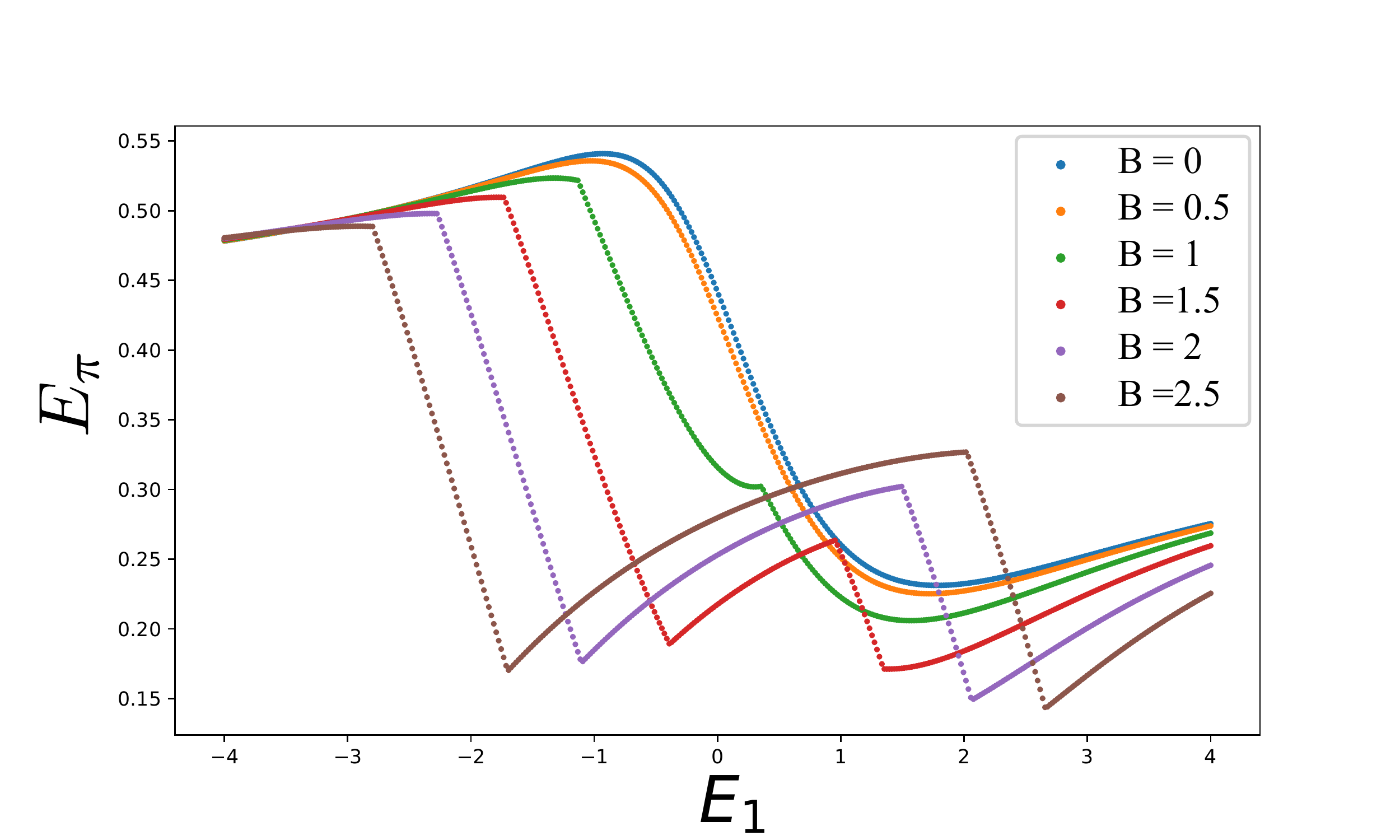}
		\caption{$E_\pi$ versus $E_1$ at several values of magnetic field.}
		\label{fig:E_T_SO_perp}
	\end{subfigure}
	\begin{subfigure}{0.45\textwidth}
		\includegraphics[width=\textwidth]
		{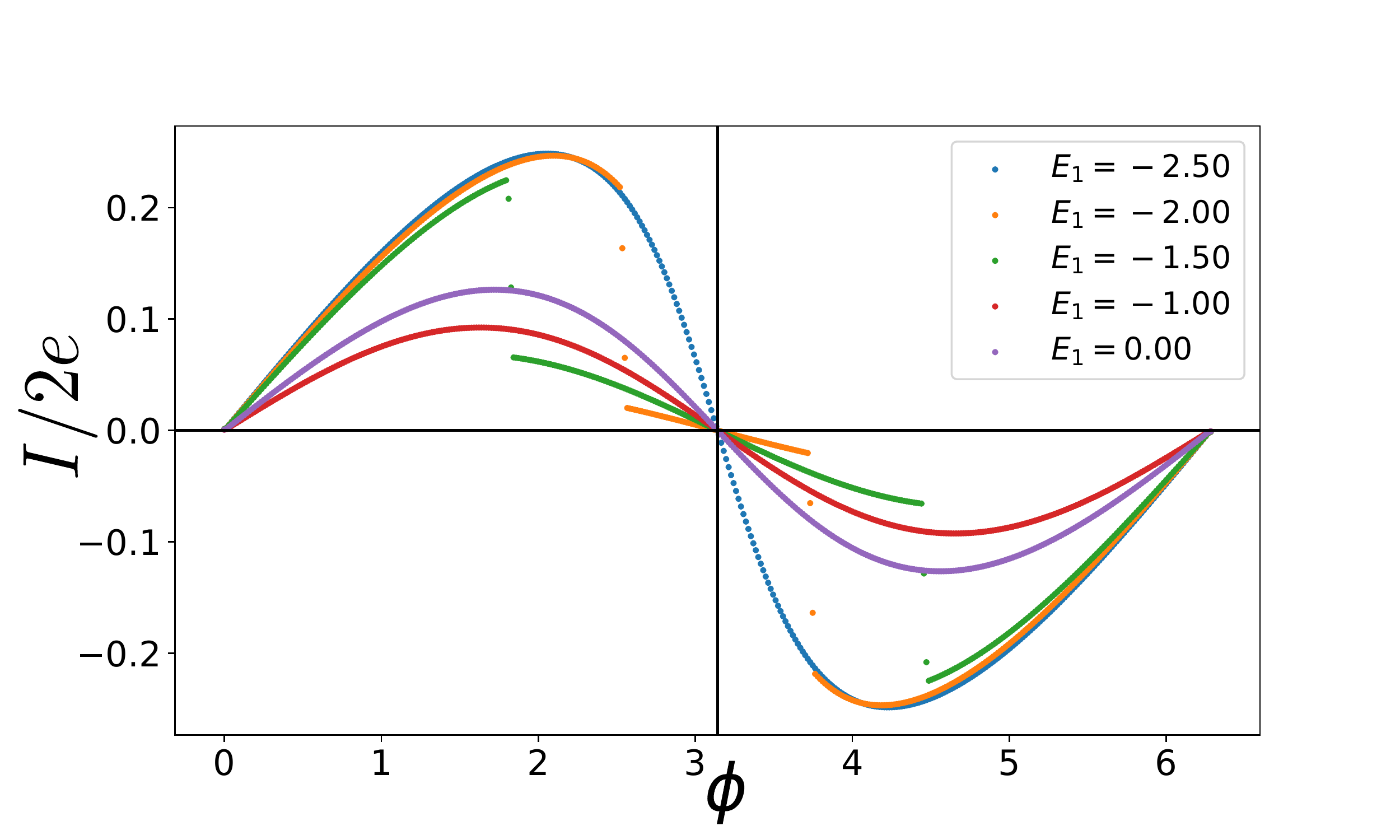}
		\caption{The phase dependence of the superconducting current at $B=2$ for several values of $E_1$.}
		\label{fig:I_SO_perp}
	\end{subfigure}
	\begin{subfigure}{0.45\textwidth}
		\includegraphics[width=\textwidth ]
		{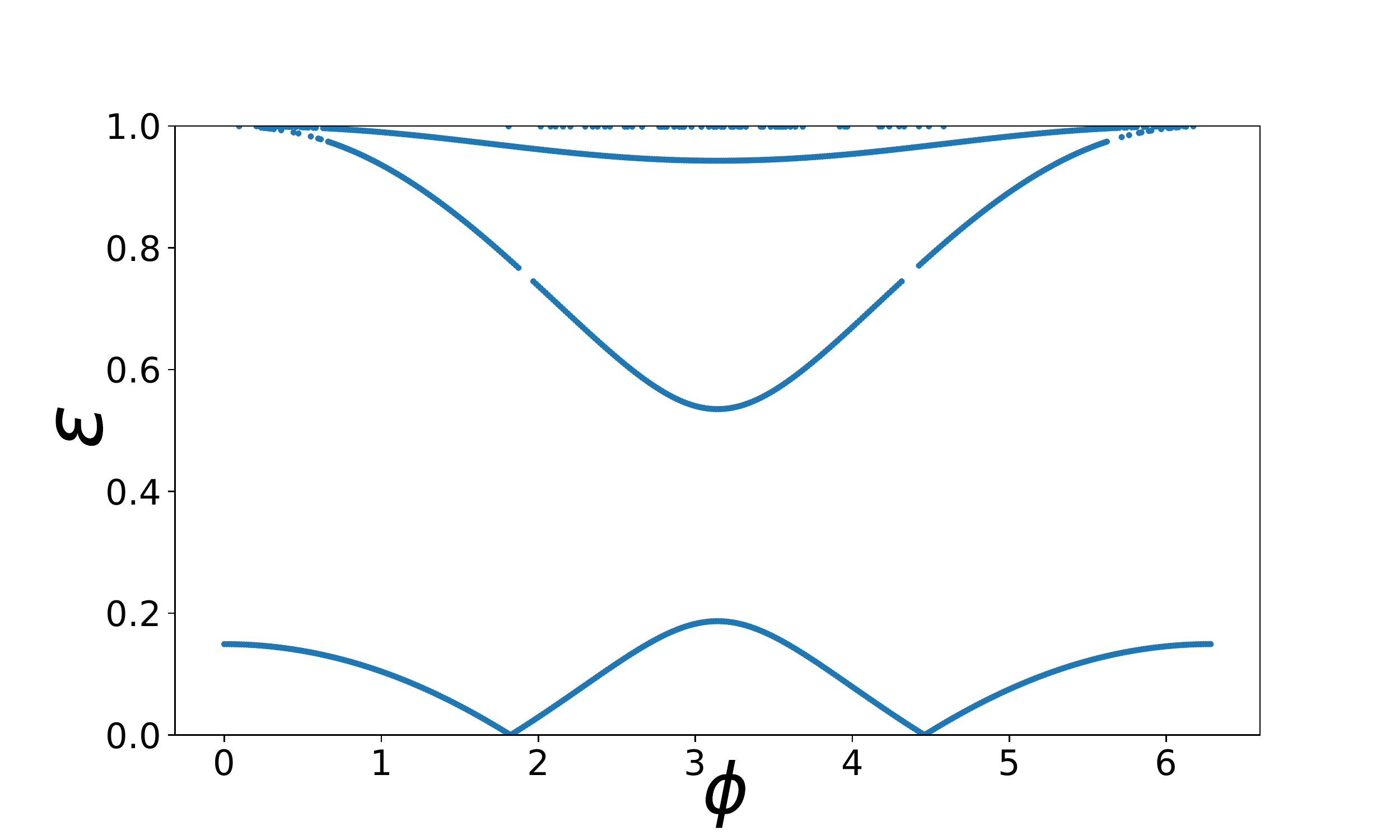}
		\caption{The phase dependence of ABS energies at $E_1=-1.5$ and $B=2$.}
		\label{fig:ABS_SO_perp}
	\end{subfigure}
	\begin{subfigure}{0.45\textwidth}
	\includegraphics[width=\textwidth ]
	{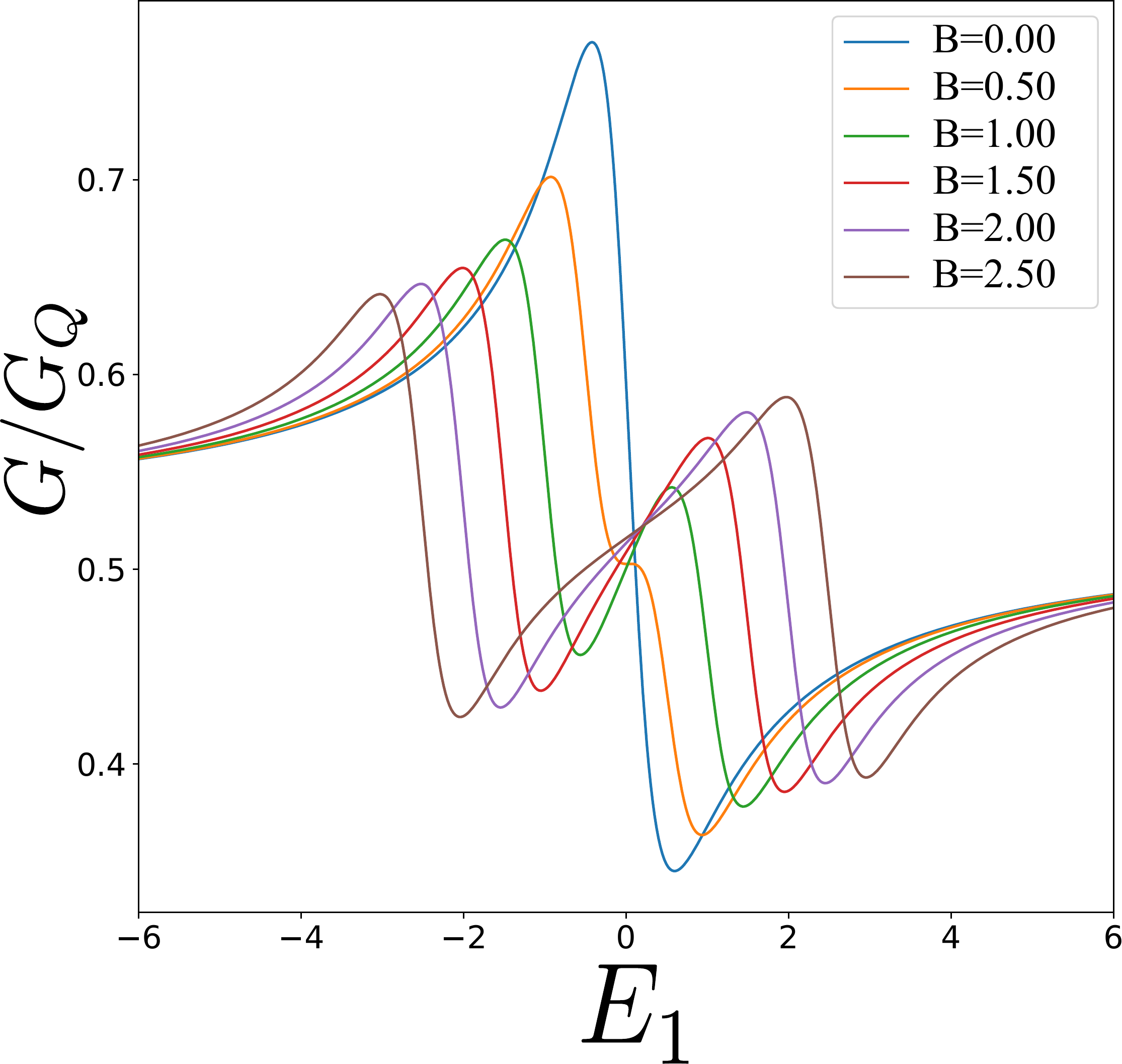}
	\caption{Normal zero-voltage conductance versus $E_1$ at several values of magnetic field.}
	\label{fig:normal_conductane_Perp_SO}
\end{subfigure}
\caption{Example C. Well-developed Fano features, moderate SO coupling. Magnetic field $ \boldsymbol{B}\perp \boldsymbol{\Gamma}$}
\label{fig:Hristo2}
\end{figure*}

\begin{figure*}
	\centering
	\begin{subfigure}{0.45\textwidth}
		\includegraphics[width=\textwidth]
		{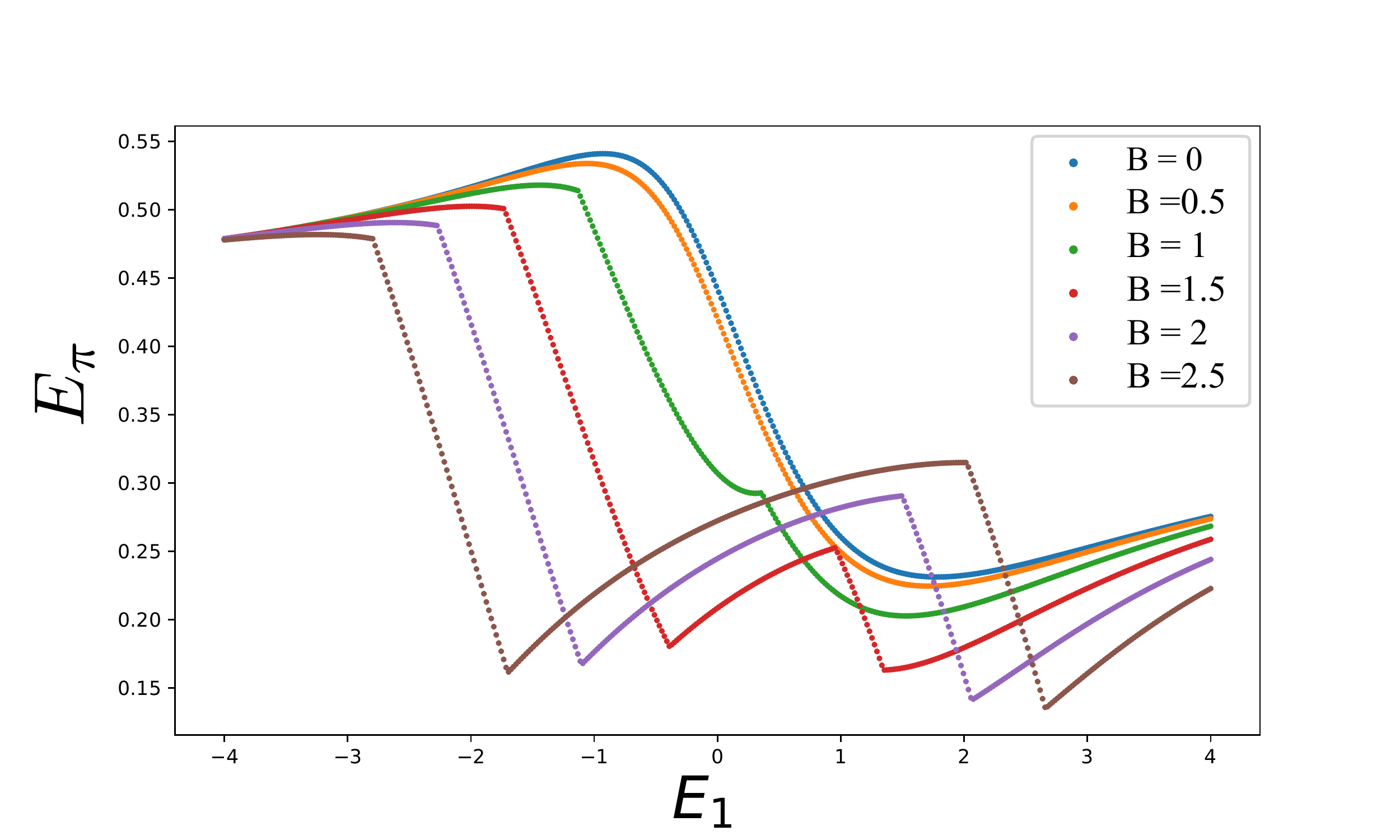}
		\caption{$E_\pi$ versus $E_1$ at several values of magnetic field.}
		\label{fig:E_T_SO_along}
	\end{subfigure}
	\begin{subfigure}{0.45\textwidth}
		\includegraphics[width=\textwidth]
		{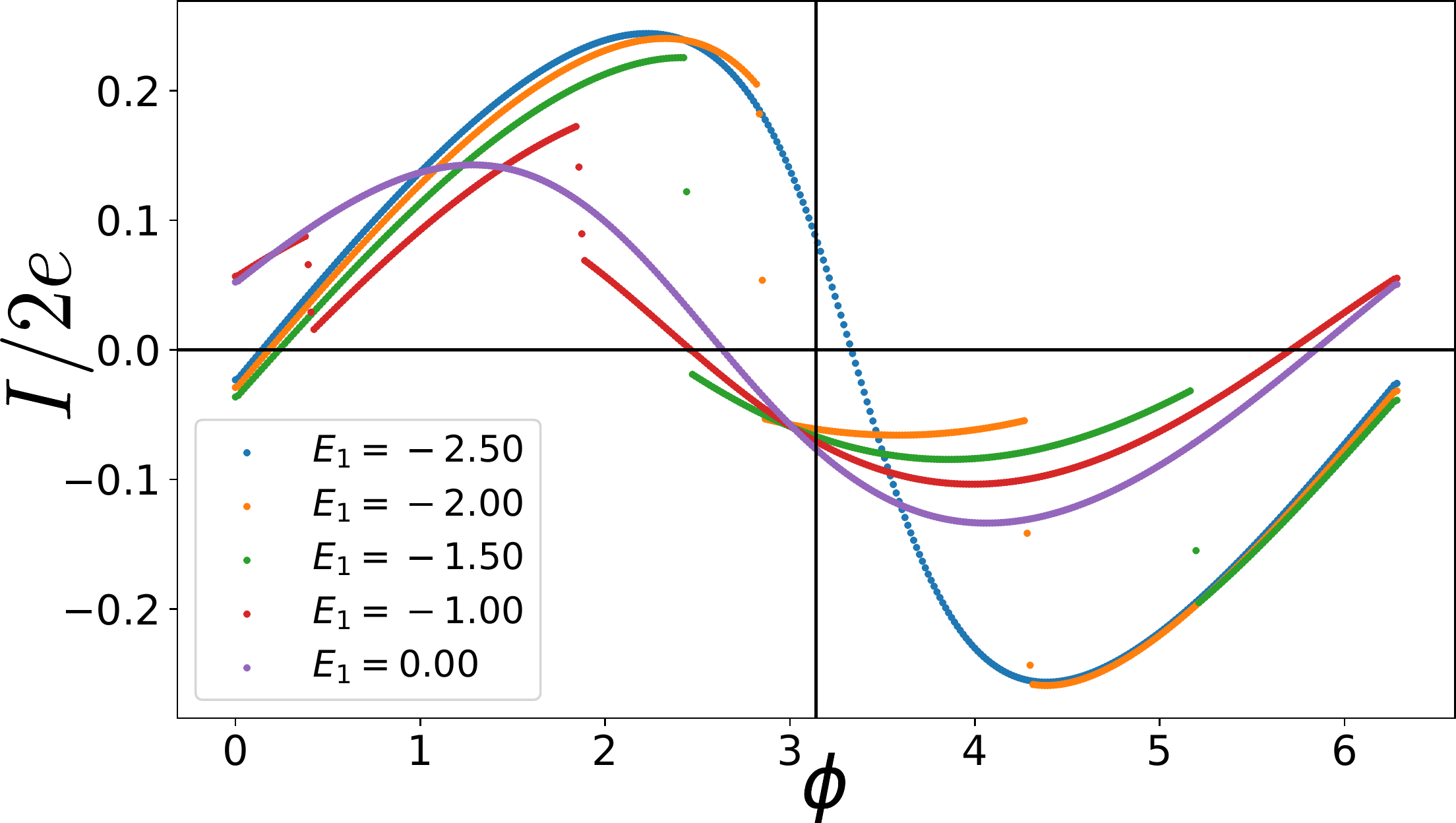}
		\caption{The phase dependence of the superconducting current at $B=1.5$ for several values of $E_1$.}
		\label{fig:I_SO_along}
	\end{subfigure}
	\begin{subfigure}{0.45\textwidth}
		\includegraphics[width=\textwidth ]
		{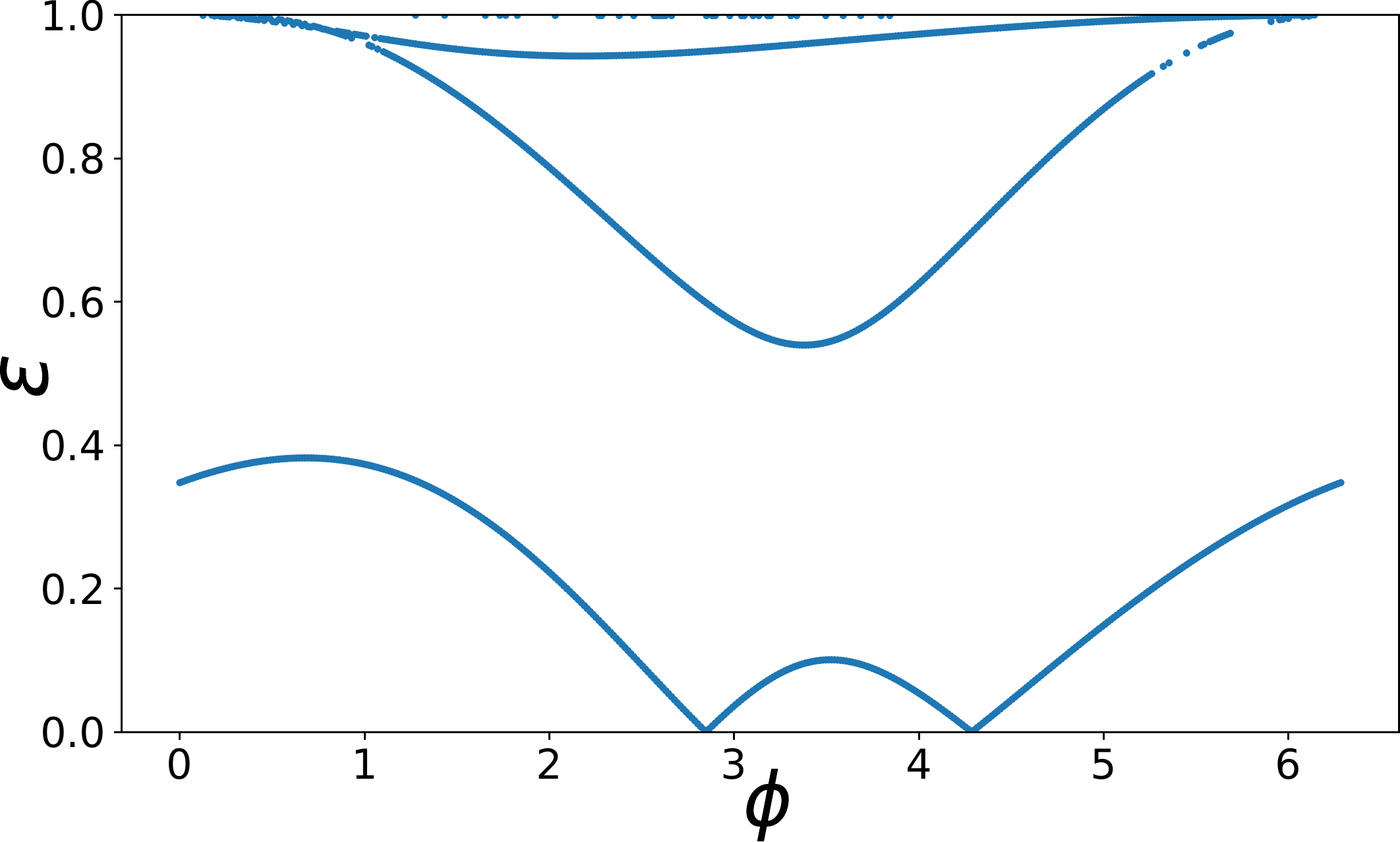}
		\caption{The phase dependence of ABS energies at $E_1=-2$ and $|B|=2$.}
		\label{fig:ABS_SO_along}
	\end{subfigure}
	\begin{subfigure}{0.45\textwidth}
	\includegraphics[width=\textwidth ]
	{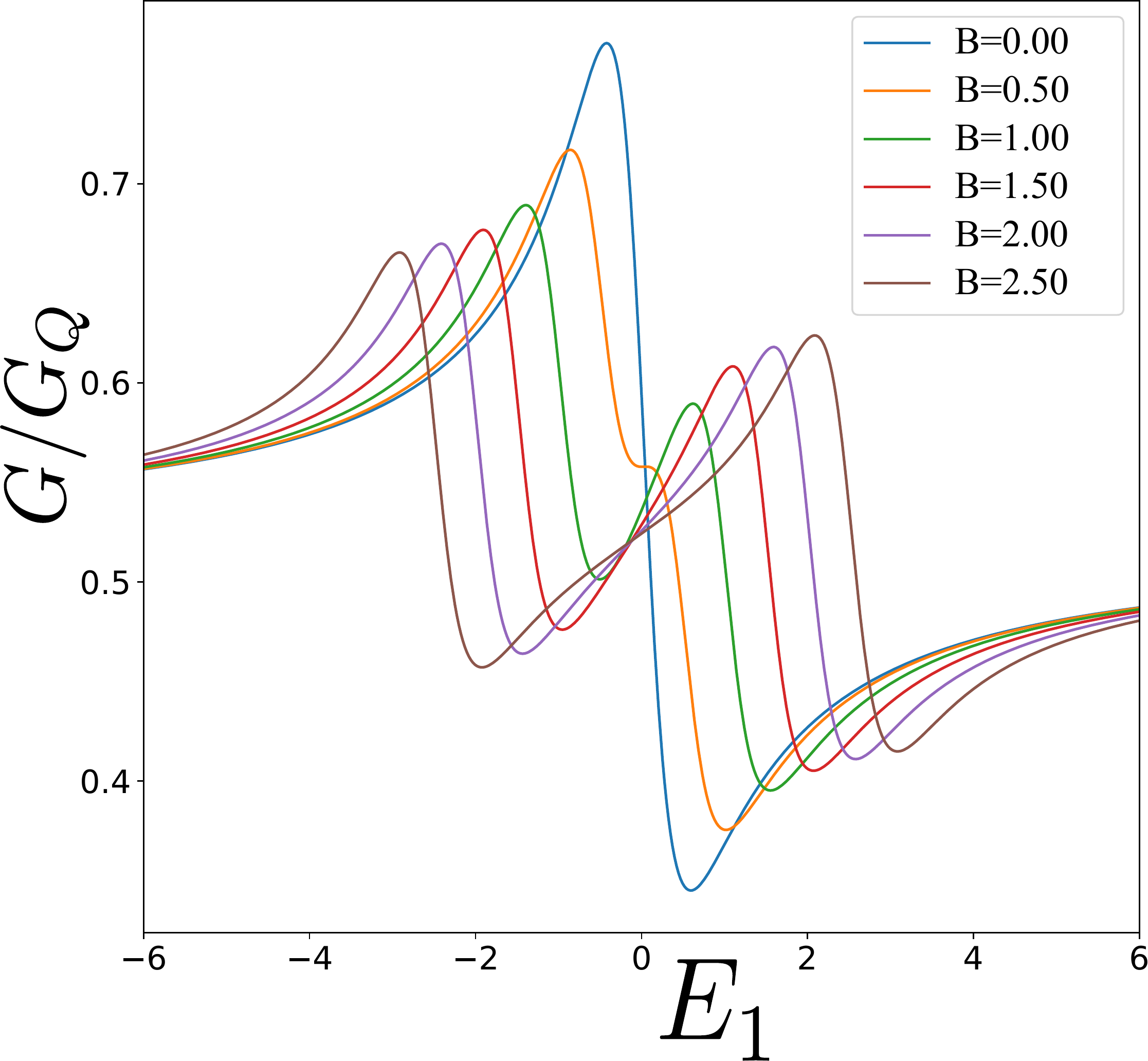}
	\caption{Normal zero-voltage conductance versus $E_1$ at several values of magnetic field.}
	\label{fig:Normal_Conductance_along}
\end{subfigure}
\caption{Example C. Well-developed Fano features, moderate SO coupling. No interaction. Magnetic field $ \boldsymbol{B} \parallel \boldsymbol{\Gamma}$. Pronounced asymmetry in $\phi$.}
\label{fig:Hristo3}
\end{figure*}

{\it Example C.} (Figs. \ref{fig:Hristo2},\ref{fig:Hristo3}) 
In this example, we illustrate the effect of SO coupling on the superconducting transport. We choose $\Gamma_2^L,\Gamma_2^R,E_2 = A(1.2,1.5,-1)$, $\kappa, \gamma_L,\gamma_R = \sqrt{A}(0.2,0.6,0.2)$, $\Gamma_1^L, \Gamma_1^R = 1.6, 3.5$. 
As to spin-dependent parameters, we choose 
$\boldsymbol{\kappa} = \sqrt{A} SO [0, 0.8, 0]$, $\boldsymbol{\gamma}_L = \sqrt{A} SO [0,0.2,0]$, $\boldsymbol{\gamma}_R = \sqrt{A} SO [0,0.1,0]$ with $SO=1$ that gives $T_0 = 0.64, \Gamma_L = 1.1, \Gamma_R = 1.38, \Gamma = 2.48$ and a significant $\boldsymbol{\Gamma} = 0.45 \boldsymbol{y}$. As we see from the Figs. \ref{fig:normal_conductane_Perp_SO},\ref{fig:Normal_Conductance_along} that give the traces of normal conductance, this set also illustrates a well-developed Fano resonance with antisymmetric features split in sufficiently high magnetic field.

We consider first $\boldsymbol{B} \perp \boldsymbol{\Gamma}$. In this case, the time-reversibility provides the symmetry $\phi \leftrightarrow - \phi$ that was present in all previous plots. Let us concentrate at the $0-\pi$ energy difference (Fig. \ref{fig:E_T_SO_perp}). The curve at zero magnetic field qualitatively follows the normal conductance. Upon increasing the magnetic field we see the multiple cusps that are already familiar from Figs. \ref{fig:Hristo0}, \ref{fig:Hristo1} and indicate the spin splitting and eventual zero crossing of ABS. The shape of the trace becomes more complex, and the minimum $E_T$ becomes smaller. However, it does not reach zero that is necessary for $0-\pi$ transition. 

The phase dependence of superconducting current at $B=2$ and various $E_1$ is presented in Fig. \ref{fig:I_SO_perp}. Most curves display pronounced discontinuities manifesting the zero crossings at corresponding phases. Except for this, the dependence is rather sinusoidal corresponding to moderate transmission. It looks like the current jumps between two sin-like curves of different amplitudes.

It is interesting to see 3 ABS in the plot presenting the phase dependence of ABS energies for $E_1 = -1.5$ and $B=2$ (Fig. \ref{fig:ABS_SO_perp}). The fourth state is either shifted over the gap edge to the continuous spectrum or is present very close to the edge so we cannot resolve it with accuracy of our numerics. The lowest state displays the familiar zero crossings corresponding to the current jumps.

When we change from perpendicular to parallel field (Fig. \ref{fig:Hristo3}) we do not see much change in normal conductance: the difference between the corresponding traces in Figs. \ref{fig:Normal_Conductance_along} and \ref{fig:normal_conductane_Perp_SO} does not exceed 10 \% . This is explained by the fact that the effect is of the second order in $\boldsymbol{\Gamma}$, $\propto \boldsymbol{\Gamma}^2/\Gamma^2$, and $|\boldsymbol{\Gamma}|/\Gamma \simeq 0.2$ is not so big. We also do not see much changes in $E_T$ traces (Fig. \ref{fig:E_T_SO_perp} versus Fig. \ref{fig:E_T_SO_along}).

The most prominent effect of SO coupling is the breaking of $\phi \leftrightarrow -\phi$ symmetry in magnetic field, the effect $\propto |\boldsymbol{\Gamma}|/\Gamma $ at $B \simeq \Gamma$. We see this in Fig. \ref{fig:I_SO_along} where the current-phase dependencies for $B=2$ are now shifted sin-like curves with jumps. The values of the shift vary from trace to trace, also in sign, and are $\simeq 0.2 - 0.3 $. In addition to the shifts of the sin-like curves, the positions of jumps are shifted non-symmetrically, these shifts are $\simeq 0 -0.5$.

Non-symmetry of the phase dependence of ABS energies is clearly seen in Fig. \ref{fig:ABS_SO_along} that is done at the same parameters as Fig. \ref{fig:ABS_SO_perp}. Also, beside shift, the energy first ABS is significantly affected by the direction of the magnetic field. A fine detail is the crossing of the second and the third ABS near $\phi \approx 1$. It may seem that in the presence of SO coupling all level crossings shall be avoided, since spin is not a good quantum number. However, since $\boldsymbol{\Gamma}$ is the only  spin vector in our model, for the particular case $\boldsymbol{B} \parallel \Gamma$ the projection of spin on $\boldsymbol{B}$ is a good quantum number and the leves of different projections may cross.

\section{Conclusions}
\label{sec:conclusions}
In conclusion, we have formulated a model that accurately describes normal and superconducting transport for a situation where a high transmission in a transport channel is accompanied by propagation through a resonant localized state. The motivation came from experimental observation of a pair of $0-\pi$ transitions separated by a small interval in the gate voltage. We have presented several examples those by no means exhaust the rich parameter space of the model. More extensive exploration of this space is required to understand if the model can explain the experimental observation. 

The data, the code and the figures can be found at https://doi.org/10.5281/zenodo.5879475.

\acknowledgments

This research was supported by the European
Research Council (ERC) under the European Union's
Horizon 2020 research and innovation programme (grant
agreement No. 694272). We are indebted to V. Levajac, J. Y. Wang, and L.P. Kouwehnoven for sharing the experimental resuls prior the publication and for numerous inspiring  discussions that initiated this research.

\bibliography{bibliography}

\end{document}